\documentclass[aps,preprint]{revtex4}
\usepackage{epsfig,graphicx,color,appendix}
\usepackage{dcolumn}
\usepackage{bm}
\newcommand{\hs}{\hspace*{0.5cm}}

\newcommand{\be}{\begin{equation}}
\newcommand{\ee}{\end{equation}}
\newcommand{\bea}{\begin{eqnarray}}
\newcommand{\eea}{\end{eqnarray}}
\newcommand{\bary}{\begin{array}}
\newcommand{\eary}{\end{array}}
\newcommand{\bit}{\begin{itemize}}
\newcommand{\eit}{\end{itemize}}
\newcommand{\ben}{\begin{enumerate}}
\newcommand{\een}{\end{enumerate}}
\newcommand{\crn}{\nonumber \\}
\newcommand{\nn}{\nonumber}

\newcommand{\al}{\alpha}
\newcommand{\la}{\lambda}
\newcommand{\bet}{\beta}
\newcommand{\ga}{\gamma}

\newcommand{\om}{\omega}

\newcommand{\fr}{\frac}

\newcommand{\bc}{\begin{center}}
\newcommand{\ec}{\end{center}}

\begin{document}


\vskip 5.5 cm

\title{The muon anomalous magnetic moment in the supersymmetric economical 3-3-1  model}

\author{ D. T. Binh\footnote{dtbinh@iop.vast.ac.vn}, D. T. Huong\footnote{dthuong@iop.vast.ac.vn} and  H. N. Long\footnote{hnlong@iop.vast.ac.vn}}

\affiliation{Institute of Physics, Vietnam Academy of Science and Technology, 10 Dao Tan, Ba Dinh,
 Hanoi,Vietnam }

\vspace*{0.5cm}


\begin{abstract}

We investigate the muon anomalous magnetic moment in the context of the supersymmetric version
of the economical 3-3-1  model. We compute the 1-loop contribution of super-partner particles.
We show that the contribution of superparticle loops become significant when $\tan \gamma $ is large.
We investigate for both small and large  values  of $\tan \gamma$. We find the region of the
 parameter space where the slepton masses are of a few hundreds  GeV is favoured by
 the muon $g-2$ for small  $\tan \gamma $  ($\tan \gamma \sim 5 $).
Numerical estimation  gives the mass of supersymmetric  particle, the mass of gauginos $m_G \sim 700$ GeV
and light  slepton mass $m_{\tilde{L}}$  is of order $ \cal O$ (100)  GeV. When $\tan{\gamma}$ is
large ($\tan{\gamma} \sim 60$), the mass of charged slepton $m_{\tilde{L}}$ and the mass
of gauginos $m_G$ $\sim$    $\cal O$(1) TeV while the mass of sneutrino  $\sim 450$ GeV
is in the reach of LHC.
\end{abstract}

\maketitle

%
%

\section{Introduction}

    The muon magnetic dipole moment (MDM), written in terms of $a_\mu=\frac{g_\mu-2}{2}$, is one of the most highly measured quantity in particle physics. The current discrepancy between experimental value and that predicted by Standard Model (SM) $\Delta a_\mu=a_\mu^{(exp)}-a_\mu^{(SM)}$ is $3.6 \sigma$ \cite{PDG}. This discrepancy demands an explanation. Efforts in both the experimental and theoretical fronts are taking to address this issue. On the theoretical front, there are models of new physics. One of these models is the class of $ SU(3)_C\times SU(3)_L\times U(1)$  (3-3-1) models  \cite{331ver1}-\cite{331verE}. In this class of models,  the $SU(2)_L$ gauge symmetry group is extended to $SU(3)_L$ and has some intriguing features: They can give of  the generation number problem; dark matter;  small neutrinos mass and their mixing and some problems of the Early Universe \cite{Lepto}. Among the  3-3-1 models, there is one version called  the economical 3-3-1 model (E331).  The economical 3-3-1 model \cite{331verE} is a model with just two Higgs
 triplet. The Higgs sector in the E331 model  is very simple and  consists of two massive neutral  Higgs scalars,  one massive charged Higgs and eight Goldstone bosons. Because of the expansion of  the gauge group,  the 3-3-1 models contain new particles such as new Higgs, new gauge bosons as well as new fermions. Due to the  interactions  of the muon with some new heavy particles, the 3-3-1 models can give new contribution to the muon MDM. The muon MDM problem is also investigated in this class of models \cite{g2muon331v1}- \cite{g2muon331v4}. However this class of models cannot
probably explain the $(g-2)_\mu$ anomaly  if the $SU(3)_L$ symmetry breaking scale is larger than ${\mathcal O}$(1) TeV \cite{queroslong,Czarnecki}.

     Supersymmetry (SUSY) is one of the most promising candidates of the new physics
beyond Standard Model. In SUSY models, the Higgs sector is very constrained and  the quadratic divergences are cancelled out and hence offer a solution to the naturalness problem \cite{SUSYnaturalness}.
   In addition, precision measurements of the gauge coupling constants strongly suggest SUSY grand unified theory \cite{SUSYUnify}.

     There are works in which the muon MDM is calculated in the framework of SUSY models \cite{MSSMg-2}. In minimal supersymmetric  standard model (MSSM) \cite{MSSM}, the SUSY contribution is proportional to
     $\tan{\beta}$ which is the ratio of the vacuum expectation values of two Higgs fields and suppression factor $\left( \frac{m_\mu}{M_{SUSY}}\right)^2$. If $\tan \beta$ is large then $a_\mu^{SUSY}$ can be as large as
 $\Delta a_\mu$ \cite{PDG}  or if low energy SUSY exits, the SUSY contribution to the muon $g-2$ can be large enough to address the muon MDM anomaly provided the masses of supersymmetry particle, order $ {\cal O}$ (100 GeV) or $ {\cal O}$(TeV), which are in the reach of the  LHC. Thus the muon MDM can be possibly originated from SUSY contributions.

It is natural to investigate the muon MDM in the SUSY  version of the
 3-3-1 models. In this work we will investigate the muon MDM in the framework of the supersymmetric
version of the economical 331 model (SUSYE331). The SUSY version of the economical 3-3-1 model  (SUSYE331) was proposed in \cite{SUSYE331}. In the SUSYE331 model  the region of parameter space can be expanded comparing to that of the MSSM. The works given in \cite{LFV1, LFV2} showed  that the interested region
of parameter space to study the LFV decay process can be expanded to the
limit of the small value of $\tan \gamma$ and the slepton mass of at least
 one generation can be taken in the  ${\cal O}$(100) GeV energy scale.

We will first derive the analytical expressions for all one-loop  contributions
coming from the SUSYE331 and show that the contribution of the
 SUSYE331 model to the muon MDM is proportional to the values of $\tan\gamma$ and inverse proportional to the  values of
the slepton masses. We will show the magnitude of
contribution to the muon MDM  for each choice of the value of the $\tan \gamma$.
Because of appearance of new particles in
the SUSYE331, we expect finding the interested region of the SUSY parameter space
in the limit of the small $\tan \gamma$.
We are particularly interested in exploring some numerical results of
the SUSYE331 contribution to the muon MDM  in the limits where 
SUSYE331 slepton and gaugino masses are of order TeV  and small values of $\tan \gamma$.

The paper is organized as follows. In section 2 we will briefly review the SUSYE331 model. In section III and IV
 we will go through the neutralinos and charginos sectors of the model. Section V and VI are devoted for
  diagonalizing the mass matrix of the smuon, sneutrino and the muon-chargino-sneutrino interaction.
In section VII we will calculate the muon MDM in the weak eigenstate. Section VIII is devoted for numerical
 calculation and bounds on masses. We will summarize our results in section IX.

\section{A review of the model}
\label{model} In this section we first recapitulate the basic
elements of the supersymmetric economical 3-3-1 model
\cite{SUSYE331}. The superfield content in this paper is defined
as follows:
 \begin{equation}
  \widehat{F}= (\widetilde{F}, F),\hs \widehat{S} =
(S, \widetilde{S}),\hs \widehat{V}= (\lambda,V),
 \end{equation}
 where the components $F$, $S$ and $V$ stand for the fermion, scalar and
vector fields while their superpartners are denoted as
$\widetilde{F}$, $\widetilde{S}$ and $\lambda$, respectively.

The superfield content of the model with ananomaly-free fermionic content transforms under the 3-3-1 gauge
group as
\[
\widehat{L}_{a L}=\left(\widehat{\nu}_{a}, \widehat{l}_{a},
\widehat{\nu}^c_{a}\right)^T_{L} \sim (1,3,-1/3),\hs
  \widehat {l}^{c}_{a L} \sim (1,1,1),
\]
\[
 \widehat Q_{1L}= \left(\widehat { u}_1,\
                        \widehat {d}_1,\
                        \widehat {u}^\prime
 \right)^T_L \sim (3,3,1/3),
 \]
\[
\widehat {u}^{c}_{1L},\ \widehat { u}^{ \prime c}_{L} \sim
(3^*,1,-2/3),\hs \widehat {d}^{c}_{1L} \sim (3^*,1,1/3 ),
\]
\[
\begin{array}{ccc}
 \widehat{Q}_{\alpha L} = \left(\widehat{d}_{\alpha},
 - \widehat{u}_{\alpha},
 \widehat{d^\prime}_{\alpha}\right)^T_{L}
 \sim (3,3^*,0), \hs \al=2,3,
\end{array}
\]
\[
\widehat{u}^{c}_{\alpha L} \sim \left(3^*,1,-2/3 \right),\hs
\widehat{d}^{c}_{\alpha L},\ \widehat{d}^{\prime c}_{\alpha L}
\sim \left(3^*,1,1/3 \right),
\]
where the values in the parentheses denote quantum numbers based
on $\left(\mbox{SU}(3)_C\right.$, $\mbox{SU}(3)_L$,
$\left.\mbox{U}(1)_X\right)$ symmetry.
$\widehat{\nu}^c_L=(\widehat{\nu}_R)^c$ and $a=1,2,3$ is a
generation index. The primes superscript on usual quark types
($u^\prime, d^\prime$ with the electric charge $q_{u'}=2/3$ and $d'$ with
$q_{d'}=-1/3$) indicate that those quarks are exotic ones.

The two superfields $\widehat{\chi}$ and $\widehat {\rho} $ are
introduced to span the scalar sector of the economical 3-3-1
model \cite{331verE}: \bea \widehat{\chi}&=& \left(
\widehat{\chi}^0_1, \widehat{\chi}^-, \widehat{\chi}^0_2
\right)^T\sim (1,3,-1/3), \crn \widehat{\rho}&=& \left(\widehat{\rho}^+_1,
\widehat{\rho}^0, \widehat{\rho}^+_2\right)^T
\sim  (1,3,2/3). \nn
\eea
To cancel the chiral anomalies of higgsino sector, two extra superfields $\widehat{\chi}^\prime$
and $\widehat {\rho}^\prime $ must be added as follows \bea
\widehat{\chi}^\prime&=& \left (\widehat{\chi}^{\prime 0}_1,
\widehat{\chi}^{\prime +},\widehat{\chi}^{\prime 0}_2
\right)^T\sim ( 1,3^*,1/3), \crn \widehat{\rho}^\prime &=& \left
(\widehat{\rho}^{\prime -}_1,
  \widehat{\rho}^{\prime 0},  \widehat{\rho}^{\prime -}_2
\right)^T\sim (1,3^*,-2/3). \nn \eea

In this model, the $ \mathrm{SU}(3)_L \otimes \mathrm{U}(1)_X$
gauge group is broken via two steps:
 \begin{equation}
  \mathrm{SU}(3)_L \otimes
\mathrm{U}(1)_X \stackrel{w,w'}{\longrightarrow}\ \mathrm{SU}(2)_L
\otimes \mathrm{U}(1)_Y\stackrel{v,v',u,u'}{\longrightarrow}
\mathrm{U}(1)_{Q},\label{stages}
 \end{equation}
 where the VEVs are defined by
\bea
 \sqrt{2} \langle\chi\rangle^T &=& \left(u, 0, w\right), \hs \sqrt{2}
 \langle\chi^\prime\rangle^T = \left(u^\prime,  0,
 w^\prime\right),\label{ct1}\\
\sqrt{2}  \langle\rho\rangle^T &=& \left( 0, v, 0 \right), \hs
\sqrt{2} \langle\rho^\prime\rangle^T = \left( 0, v^\prime,  0
\right).\nn\eea
The VEVs $w$ and $w^\prime$ are responsible for
the first step of the symmetry breaking while $u,\,  u^\prime$ and
$v,\, v^\prime$ are for the second one. Therefore, they have to
satisfy the constraints:
 \be
 u,\ u^\prime,\ v,\ v^\prime
\ll w,\ w^\prime. \label{contraint}
 \ee

The vector superfields $\widehat{V}_c$, $\widehat{V}$ and
$\widehat{V}^\prime$ containing the usual gauge bosons are,
respectively, associated with the $\mathrm{SU}(3)_C$,
$\mathrm{SU}(3)_L$ and $\mathrm{U}(1)_X $ group factors. The
colour and flavour vector superfields have expansions in the
Gell-Mann matrix bases $T^a=\lambda^a/2$ $(a=1,2,...,8)$ as
follows
\bea \widehat{V}_c &=& \fr{1}{2}\lambda^a
\widehat{V}_{ca},
\widehat{\overline{V}}_c=-\fr{1}{2}\lambda^{a*}
\widehat{V}_{ca}; \widehat{V} = \fr{1}{2}\lambda^a
\widehat{V}_{a}, \widehat{\overline{V}}=-\fr{1}{2}\lambda^{a*}
\widehat{V}_{a},\nn\eea
where the overbar $^-$ indicates complex
conjugation. The vector superfields associated with
$\mathrm{U}(1)_X$ are normalized as follows
 \[ X \hat{V}'= (XT^9)
\hat{B}, \hs T^9\equiv\fr{1}{\sqrt{6}}\mathrm{diag}(1,1,1).\]
 The gluons are denoted by $g^a$ and their gluino partners
by $\lambda^a_{c}$, with $a=1, \ldots,8$. In the electroweak
sector, $V^a$ and $B$ stand for the $\mathrm{SU}(3)_{L}$ and
$\mathrm{U}(1)_{X}$ gauge bosons with their gaugino partners
$\lambda^a_{V}$ and $\lambda_{B}$, respectively.

With the given superfields, the full Lagrangian is defined by
$\mathcal{L}_{susy}+\mathcal{L}_{soft}$ where the first term is
supersymmetric part, whereas the last term breaks the
supersymmetry explicitly. The interested reader can find more
details about the Lagrangian in \cite{SUSYE331}. In
the followings, we only interest in terms relevant to our calculations.

\section{The neutralinos and chargino sectors}
\label{neutral}
In the SUSYE331 the neutralinos are mixed by $11 \times 11$ matrix and the charginos
are mixed by $5\times 5$ matrix. It is difficult to find the exactly mass eigenstate
of these mixing mass matrices. Hence, we have to find the approximation method.

\subsection{The neutralino sector}
 The neutralino mass terms are given in \cite{gaugino} as follows:
  \begin{equation}
  \mathcal{L}=\left(\widetilde{\psi^{o} }\right)^\dag
  M_{\widetilde{N}}\widetilde{\psi^o},
    \end{equation}
with
\bea
\widetilde{\psi}^o= \left( \widetilde{\chi^o_1 }
  ,  \widetilde{\chi^{o\prime}_1 }, \widetilde{\chi^o_2 }
  , \widetilde{\chi^{o\prime}_2 }, \widetilde{\rho^o_1},
  \widetilde{\rho^{o\prime}_1}, \lambda_B,
  \lambda_3,
 \lambda_8, \right.  
\left.    \lambda_X=\frac{\lambda_4
  +i\lambda_5}{2},
  \lambda_{X^*}=\frac{\lambda_4
  -i\lambda_5}{2}\right)
\nn  \eea
and the mass matrix $M_{\widetilde{N}}$ has the form as follows
\begin{equation}
M_{\widetilde{N}}=
\left(
 \begin{array}{ccccccccccc}
0 & -\mu_\chi & 0 & 0 & 0 & 0 & -\frac{g^\prime u}{3\sqrt{6}}
 & \frac{gu}{2} & \frac{g u}{2\sqrt{3}} & \frac{gw}{\sqrt{2}} & 0 \\
 -\mu_\chi  & 0 & 0& 0& 0 & 0 & \frac{g^\prime u^\prime}{3\sqrt{6}}
 & \frac{gu^\prime}{2}& \frac{g u^\prime}{2\sqrt{3}} & \frac{gw^\prime}
 {\sqrt{2}} & 0 \\
 0 & 0 & 0 &  &-\mu_\chi & 0  &- \frac{g^\prime w}{3\sqrt{6}}
  & 0& -\frac{gw}{\sqrt{3}} & 0&
  \frac{gu}{\sqrt{2}}\\
  0& 0& -\mu_\chi  &0 & 0 & 0 & \frac{g^\prime w^\prime}{3\sqrt{6}}
  & 0 & -\frac{gw^\prime}{\sqrt{3}} & 0 & \frac{gu^\prime}{\sqrt{2}} \\
 0 & 0 & 0 & 0 & 0 & -\mu_\rho & \frac{2g^\prime v}{3\sqrt{6}} &
 -\frac{gv}{2} & \frac{gv}{2\sqrt{3}} & 0 & 0 \\
 0 & 0 & 0 & 0 & -\mu_\rho & 0 & -\frac{2g^\prime v^\prime}{3\sqrt{6}}
 & -\frac{gv^\prime}{2} & \frac{gv^\prime}{2\sqrt{3}} & 0& 0 \\
 -\frac{g^\prime u}{3\sqrt{6}}& \frac{g^\prime u^\prime}{3\sqrt{6}} &
 -\frac{g^\prime w}{3\sqrt{6}}
 &  \frac{g^\prime w^\prime}{3\sqrt{6}} &  \frac{2g^\prime v}{3\sqrt{6}} &
 -\frac{2g^\prime v^\prime}{3\sqrt{6}} & m_B & 0 & 0 & 0 & 0 \\
  \frac{gu}{2} & \frac{gu^\prime}{2} & 0 & 0& -\frac{gv}{2}
  & -\frac{gv^\prime}{2} & 0& m_{\lambda_3} & 0 & 0 & 0 \\
\frac{gu}{2\sqrt{3}} & \frac{gu^\prime}{2\sqrt{3}} &
-\frac{gw}{\sqrt{3}} & -\frac{gw^\prime}{\sqrt{3}} &
\frac{gv}{2\sqrt{3}}  &  \frac{gv^\prime}{2\sqrt{3}}
 & 0 & 0& m_{\lambda_8}& 0 & 0 \\
\frac{gw}{2} & \frac{gw^\prime}{2} & 0 & 0 & 0 & 0 & 0 & 0
& 0 & m_{\lambda_{45}} & 0 \\
0 & 0 & \frac{gu}{2} & \frac{gu^\prime}{2} & 0 & 0 & 0& 0 & 0 & 0
& m_{\lambda_{45}}
        \end{array}
                  \right),\crn
                 \label{an4}
 \end{equation}
 where $m_{\lambda_4} = m_{\lambda_5} \equiv
 m_{\lambda_{45}}$.

 In general, we can find a new basis in which the mass matrix $M_{\widetilde{N}}$ has the diagonal
 form by finding an unitary matrix $N$ satisfied
\bea
 N^* M_{\widetilde{N}}N^\dag &=& Diag
 (m^2_{\widetilde{\chi}_1}, m^2_{\widetilde{\chi}_2}, m^2_{\widetilde{\chi}_3},
 m^2_{\widetilde{\chi}_4}, m^2_{\widetilde{\chi}_5}, m^2_{\widetilde{\chi}_6},
 m^2_{\widetilde{\chi}_7},  m^2_{\widetilde{\chi}_8}, m^2_{\widetilde{\chi}_9},
 m^2_{\widetilde{\chi}_{10}}, m^2_{\widetilde{\chi}_{11}}).
\nn  \eea
As have done in \cite{gaugino}, we assume that
   \bea
   v,v^\prime, u, u^\prime, w, w^\prime    &\ll&     \left|\mu_{\rho}
   -m_B
   \right|,\left|\mu_{\rho}-m_{\lambda_3}
   \right|,\left|\mu_{\rho}-m_{\lambda_8 }
   \right|, \left|\mu_{\rho}-m_{\lambda_{45}} \right|
\label{an55}
   \eea
and
   \bea
   v,v^\prime, u, u^\prime, w, w^\prime &\ll& \left|\mu_{\chi}-m_B
   \right|,\left|\mu_{\chi}-m_{\lambda_3}
   \right|,\left|\mu_{\chi}-m_{\lambda_8}
   \right|, \left|\mu_{\chi}-m_{\lambda_{45}} \right|. \label{an5}
   \eea

   In these limits, we get the neutralino mass eigenstates by using perturbation.
    At the first order of the perturbation, we obtain the
    mixing matrix $N$ such as:
\bea
  N=\left(%
    \begin{array}{ccccccccccc}
      0 & 0 & 0 & 0 & 0 & 0 & 1 & 0 & 0 & 0 & 0 \\
      0 & 0 & 0 & 0 & 0 & 0 & 0 & 1 & 0 & 0 & 0 \\
      0 & 0 & 0 &0 & 0 & 0 & 0 & 0 & 1 & 0 & 0\\
     0 & 0 & 0 &0 & 0 & 0 & 0 & 0 & 0 & 1 & 0\\
      0 & 0 & 0 & 0 & 0 & 0 & 0 & 0 & 0 & 0 & 1 \\
      0 & 0 & 0 & 0 & \frac{1}{\sqrt{2}} & \frac{1}{\sqrt{2}} & 0 & 0 & 0 & 0 & 0\\
      0 & 0 & 0 &0 & \frac{1}{\sqrt{2}} & -\frac{1}{\sqrt{2}} & 0 & 0 & 0 & 0 & 0\\
       \frac{1}{\sqrt{2}} & \frac{1}{\sqrt{2}} 0 & 0 & 0 &0 & 0 & 0 & 0 & 0 & 0\\
       \frac{1}{\sqrt{2}} & -\frac{1}{\sqrt{2}} 0 & 0 & 0 &0 & 0 & 0 & 0 & 0 & 0\\
        0 & 0 & \frac{1}{\sqrt{2}} & \frac{1}{\sqrt{2}}&0 &0 & 0 & 0 & 0 & 0 & 0\\
       0 & 0 & \frac{1}{\sqrt{2}} & -\frac{1}{\sqrt{2}}&0 &0 & 0 & 0 & 0 & 0 & 0 \\
    \end{array}%
    \right),
\nn   \eea
and the mass eigenvalues as studied in \cite{gaugino}.

\subsection{Chargino sector}\label{chargino} 
In the SUSYE331 model, there are four charged gauginos and six charged Higgsinos.  In the
 basis $\psi^{+}=(\lambda_{\widetilde{\mathcal{ W}}^+}$, $\lambda_{\widetilde{\mathcal{Y}}^+}$,
$\widetilde{\rho_1}^+$, $ \widetilde{\rho_2}^+$, $\widetilde{\chi}^{\prime +})$,
$\psi^-=(\lambda_{\widetilde{\mathcal{ W}}^-}$, $\lambda_{\widetilde{\mathcal{Y}}^-}$,
$\widetilde{\rho_1}^{\prime -}$,
$\widetilde{\rho_2}^{\prime -}$, $\widetilde{\chi}^{-})$, the  Lagrangian describes  the chargino
mass terms is given  as follows
 \[ \mathcal{L}_{chargino mass} = \left(\widetilde{\psi}^{\pm}\right)^+
 M_{\widetilde{\psi}} \widetilde{\psi}^{\pm} + H.c,\]
 with the $M_{\widetilde{\psi}}$
 \[
 M_{\widetilde{\psi}} = \left(%
\begin{array}{cc}
  0 & \mathcal{M }\\
  \mathcal{M}^T & 0 \\
\end{array}%
\right),
 \]
 where $\mathcal{M}$ is $5 \times 5$ matrix  given by
 \[
 \mathcal{M}=\left(%
\begin{array}{ccccc}
  m_{\lambda_W} & 0 &\frac{gv^\prime}{\sqrt{2}} & 0 & \frac{gu}{\sqrt{2}}\\
  0 &  m_{\lambda_Y} & 0&\frac{gv^\prime}{\sqrt{2}} &\frac{gw}{\sqrt{2}} \\
  \frac{gv}{\sqrt{2}} & 0 & \mu_\rho & 0 & 0\\
  0 & \frac{gv}{\sqrt{2}} & 0 & \mu_{\rho} & 0 \\
  \frac{gu^\prime}{\sqrt{2}} & \frac{gw^\prime}{\sqrt{2}} & 0 & 0 & \mu_\chi\\
\end{array}%
\right).
 \]
In order to diagonalize charginos mass matrix, we have to find two unitary $5 \times 5$ matrices $U$ and
$V$ such that they satisfy
 \[
 U^{\ast}\mathcal{M }V^{\dag}=\left(%
\begin{array}{ccccc}
  m_{\lambda_W} & 0 & 0&0 & 0 \\
  0 &  m_2 & 0 & 0 & 0 \\
  0 & 0 &  \mu_\rho & 0 & 0 \\
  0 & 0 & 0 &  m_3 &0 \\
  0 & 0 & 0 & 0 &  \mu_\rho \\
\end{array}%
\right) \equiv  \textrm{Diag}(m_{\chi_1}^\pm, m_{\chi_2}^\pm, m_{\chi_3}^\pm, m_{\chi_4}^\pm, m_{\chi_5}^\pm),
 \]
 where
  \bea
 m_2&=&\frac{1}{4}\left[2(\mu_\chi^2+m_{\lambda_Y}^2)+g^2(w^2+w^{\prime
 2})\right.\crn &&-\left. \sqrt{(2(\mu_\chi +m_{\lambda_Y})^2+g^2(w-
 w^\prime)^2)\left(2(\mu_\chi-m_{\lambda_Y})^2+g^2(w+w^{\prime })^2\right)} \right],
 \crn
  m_3&=&\frac{1}{4}\left[2(\mu_\chi^2+m_{\lambda_Y}^2)+g^2(w^2+w^{\prime
 2})\right. \crn
 &&+\left. \sqrt{(2(\mu_\chi +m_{\lambda_Y})^2+g^2(w-
 w^\prime)^2)\left(2(\mu_\chi-m_{\lambda_Y})^2
 +g^2(w+w^{\prime })^2\right)} \right]. \nn\eea

 In order to find the matrices $U, V$ we  have to diagonalize the matrix $M^\dag M$
  by finding the matrix $C$ such that it satisfies $C^\dag M ^\dag M C\equiv M_D^2$, $M_D$
 has a diagonal form. Meanwhile, the shape of the matrices $U$ and $V$ are represented
through the matrices  $U, V$ as follows
 \[ V^\dag=C, \hs \hs  U^\star = M_D^{-1}C^\dag M^\dag.\]
In  the leading order $m_{\lambda_W},m_{\lambda_Y}, \mu_\rho,
\mu_\chi, w,
w^\prime \gg u, u^\prime v, v^\prime$,  the matrix $C$ has a form such as \bea C =\left(%
\begin{array}{ccccc}
  1 & 0 & 0 & 0 & 0 \\
 0 & \frac{A}{\sqrt{1+A^2}} & 0  &  \frac{B}{\sqrt{1+B^2}} & 0 \\
  0 & 0 & 1 & 0 & 0 \\
  0 & 0 & 0 & 0 & 1 \\
  0 &  \frac{1}{\sqrt{1+A^2}} & 0 &  \frac{1}{\sqrt{1+B^2}} & 0 \\
\end{array}%
\right),\nn \eea and the form of the matrix $U^*$ is
\bea
   U^*=\left(%
       \begin{array}{ccccc}
         1 & 0 & \frac{gv}{\sqrt{2}m_{\lambda_W}}& 0 & \frac{g u^\prime}{\sqrt{2}m_{\lambda_W}} \\
         \frac{gu}{\sqrt{2(1+A^2)^2}m_2}&\frac{2Am_{\lambda_Y}+\sqrt{2}gw}{2\sqrt{(1+A^2)}m_2}
         & 0 & \frac{gvA}{\sqrt{2(1+A^2)}m_2} &
         \frac{2\mu_\chi+\sqrt{2}Agw^\prime}{2\sqrt{(1+A^2)}m_2}   \\
         \frac{gv^\prime}{\sqrt{2} \mu_\rho} & 0 & 1 & 0 & 0 \\
         \frac{gu}{\sqrt{2(1+B^2)}m_3} & \frac{2Bm_{\lambda_Y}
         +\sqrt{2}gw}{2\sqrt{(1+B^2)}m_3}
         & 0 & \frac{Bgv}{\sqrt{2(1+B^2)}m_3} & \frac{2\mu_\chi+\sqrt{2}gw^\prime B}{\sqrt{2(1+B^2)}m_3} \\
         0 & \frac{g v^\prime}{\sqrt{2} \mu_\rho} & 0 & 1 & 0 \\
       \end{array}%
       \right)
\nn \eea
where
 \bea A&=&\frac{1}{{2\sqrt{2}(m_{\lambda_Y}w+\mu_\chi w^\prime)}}\left(  2(
\mu_\chi^2-m_{\lambda_Y}^2)+g^2(w^2-w^{\prime2}) \right. \crn
 &+& \left. \sqrt{-4(-2\mu_\chi m_{\lambda_Y}
+g^2ww^\prime )^2+(2(\mu_\chi^2+m_{\lambda_Y}^2)+g^2(w^2+w^{\prime
2}))^2} \right) , \crn
B&=&\frac{1}{2\sqrt{2}(m_{\lambda_Y}w+\mu_\chi w^\prime)} \left( -2(
\mu_\chi^2-m_{\lambda_Y}^2)-g^2(w^2-w^{\prime2}) \right. \crn
&+& \left.  \sqrt{-4(-2\mu_\chi m_{\lambda_Y}+g^2ww^\prime )^2+(2(\mu_\chi^2+m_{\lambda_Y}^2)+g^2(w^2+w^{\prime
2}))^2} \right) .\nn
 \eea

\section{Smuon  and sneutrino masses}
The superpotential  of the  model under consideration  relevant to the contribution of $(g-2)_\mu$ is
given as follows:
\bea
W^\prime&=& \mu_{0a}\hat{L}_{aL} \hat{ \chi}^{\prime}+ \mu_{ \chi} \hat{ \chi} \hat{
\chi}^{\prime}+
 \mu_{ \rho} \hat{ \rho} \hat{ \rho}^{\prime}
 +\ga_{ab} \hat{L}_{aL} \hat{ \rho}^{\prime} \hat{l}^{c}_{bL} + \la_{a} \epsilon \hat{L}_{aL}
\hat{\chi} \hat{\rho}+ \la^\prime_{ab} \epsilon \hat{L}_{aL} \hat{L}_{bL} \hat{\rho},
\nn \eea
with $\mu_{0a}, \mu_{\rho}$ and $\mu_{\chi}$ have mass dimension, the other coefficients in
$W^\prime$ are dimensionless and $\la^\prime_{ab}= - \la^\prime_{ba}$.

Relevant soft breaking terms are  obtained by
\bea -\mathcal{L}_{SMT}&=&
M^2_{ab}\widetilde{L}_{aL}^\dagger \widetilde{L}_{bL}
+ m^2_{ab}\widetilde{l}_{aL}^{c *}
\widetilde{l}_{bL}^c   + \left\{ M^{\prime 2}_a \chi^\dagger\widetilde{L}_{aL}
+ \eta_{ab}\widetilde{L}_{aL}\rho ^\prime \widetilde{l}_{Lb}^c + \upsilon_{a}\epsilon\widetilde{L}_{aL}\chi\rho
\right. \crn
&+& \left.  \varepsilon_{ab}\epsilon
\widetilde{L}_{aL}\widetilde{L}_{bL}\rho  \right. 
+ \left. \om_{a\alpha j }\widetilde{L}_{a
L}\widetilde{Q}_{\alpha L}\widetilde{d}_{j L}^c
+ \om^\prime_{a\alpha \beta }\widetilde{L}_{a
L}\widetilde{Q}_{\alpha L}\widetilde{d^\prime}_{\beta L}^c
+ H.c.  \right\},  \label{mme}
\eea
 where
$\varepsilon_{ab}= - \varepsilon_{ba}$. This Lagrangian is also responsible for sfermion masses.
The sfermion masses are obtained by combining of the soft terms, D-terms and F-terms. The
interested reader can see in \cite{chargedlepton}.  In general, there are flavour mixing in
slepton mass matrix. However, the large flavour mixing in slepton sector can create a mismatch for
the lepton flavor decay processes of muon and tauon \cite{LFV1, LFV2}. In this work,  we
assume that the flavour mixing matrix elements are not so large. We can ignore all the flavour mixing
terms. The mass matrix for smuon can be written as
\bea M_{smuon}= \left(%
\begin{array}{cc}
  m^2_{\widetilde{\mu} L} & m^2_{\widetilde{\mu} LR} \\
  m^2_{\widetilde{\mu} LR} & m^2_{\widetilde{\mu} R}\\
\end{array}%
\right), \label{chag} \eea
where
 \bea m^2_{\widetilde{\mu} L} &=&M^2_{22} + \fr 1 4 \mu_{02}^2
 + \fr{g^2}{2} \left(- H_3
 + \fr{1}{\sqrt{3}}H_8 - \fr{2 t^2}{3}H_1 \right) +
\fr{
 v^{\prime 2}}{18}\ga_{22}^2+\fr{1}{18}\la_2^2
(u^2+w^2), \\ m^2_{\widetilde{\mu} R}
&=& \left(m^2_{22} + \fr{ v^{\prime 2}}{18}\ga_{22}^2 +  g^2 t^2 H_1 \right), \\
m^2_{\widetilde{\mu} LR} &=&\frac{1}{\sqrt{2}}\left(\eta_{22} v^\prime + \fr 1 6 \mu_\rho
\ga_{2}v\right),\nn
 \eea
 with
 \bea
H_3 & = & -\fr 1 4 \left(u^2 \fr{\cos 2\bet}{s_\bet^2} + v^2 \fr{\cos 2\ga}{c_\ga^2} \right),
\label{sf184}\\ H_8  & = & \fr{1}{4\sqrt{3}} \left[ v^2 \fr{\cos 2\ga}{c_\ga^2}- (u^2 - 2w^2)
\fr{\cos 2\bet}{s_\bet^2} \right],\label{sf185} \\
H_4 &=&-\fr 1 2 u w \fr{\cos 2\beta}{s^2_\beta}\label{sf199}, \\ H_1  & = &  \fr 1 6 \left[(u^2 +w^2)
\fr{\cos 2\bet}{s_\bet^2} + 2 v^2 \fr{\cos 2\ga}{c_\ga^2} \right], \label{sf194}\eea
and $ \tan \bet = \fr{u}{u^\prime} = \fr{w}{w^\prime}, \hs \tan \ga = \fr{v^\prime}{v},
\hs s_\beta= \sin \beta, \hs  c_\beta=\cos\beta$.

Diagonalizing the mass matrix given in Eq.(\ref{chag}), we can obtain the mass eigenvalues as
follows:
\bea m_{\widetilde{\mu}_L}^2 &=&\frac{1}{2}\left( m^2_{\widetilde{\mu} L}+m^2_{\widetilde{\mu}
R}-\Delta\right), \\ m_{\widetilde{\mu}_R}^2 &=&\frac{1}{2}\left( m^2_{\widetilde{\mu}
L}+m^2_{\widetilde{\mu} R}+\Delta\right),\eea where $\Delta =
\sqrt{\left(m^2_{\widetilde{\mu} L}-m^2_{\widetilde{\mu} R}
\right)^2+4m^4_{\widetilde{\mu} LR}}$.

 The mass eigenstates are  given,  respectively
 \bea
\left(%
\begin{array}{c}
   \widetilde{\mu}_L \\
   \widetilde{\mu}_R \\
\end{array}%
\right) = \left(%
\begin{array}{cc}
 s_{\theta_{\widetilde{\mu}}} & -c_{\theta_{\widetilde{\mu}}} \\
 c_{\theta_{\widetilde{\mu}}} & s_{\theta_{\widetilde{\mu}}} \\
\end{array}%
\right) \left(%
\begin{array}{c}
 l_{\widetilde{\mu}_R} \\
  l_{\widetilde{\mu}_L} \\
\end{array}%
\right) \equiv U_{\widetilde{\mu}}^{-1}\left(%
\begin{array}{c}
 l_{\widetilde{\mu}_R} \\
  l_{\widetilde{\mu}_L} \\
\end{array}%
\right)\nn\eea
\bea \widetilde{\mu}_L &=& c_{\theta_{\widetilde{\mu}}}
l_{\widetilde{\mu}_R}-s_{\theta_{\widetilde{\mu}}} l_{\widetilde{\mu}_L},\\
     \widetilde{\mu}_R &=&s_{\theta_{\widetilde{\mu}}}
     l_{\widetilde{\mu}_R}+c_{\theta_{\widetilde{\mu}}}
     l_{\widetilde{\mu}_L},
\nn \eea with $s_{\theta_{\widetilde{\mu}}}=\sin
\theta_{\widetilde{\mu}}$, $c_{\theta_{\widetilde{\mu}}}=\cos
\theta_{\widetilde{\mu}}$ and the $\theta_{\widetilde{\mu}}$ is
defined through $\tan 2{\theta_{\widetilde{\mu}}}$ as follows:
$\tan 2 {\theta_{\widetilde{\mu}}} = t_{
2{\theta_{\widetilde{\mu}}}}=
\frac{2m^2_{\widetilde{\mu}LR}}{m^2_{\widetilde{\mu}L}-m^2_{\widetilde{\mu}R}}$.

 Next, We will study the muon sneutrino mass. If we ignore  mixing among sneutrinos of two first
generations, the mass of the sneutrino $m_{\widetilde{\nu}_\mu}$ has the form
 \bea
 m^2_{\tilde{\nu}_{\mu L}}& =& M^2_{22} + \fr 1 4 \mu_{02}^2 +
\fr{g^2}{2} \left( H_3
 + \fr{1}{\sqrt{3}}H_8 - \fr{2 t^2}{3}H_1 \right)
 +\fr{1}{18}v^2(\la_2^2+4
\la'^2_{c2})+ \fr{1}{18}\la_2^2 w ^2,
 \label{sf39n}\\
m^2_{\tilde{\nu}_{\mu R}} &=& M^2_{22} + \fr 1 4 \mu_{02}^2  - g^2 \left( \fr{1}{\sqrt{3}}H_8 +
\fr{ t^2}{3}H_1 \right)
 +\fr{1}{18}v^2(\la_2^2+4
\la'^2_{c2})+ \fr{1}{18}\la_2^2 u ^2.\label{sf41n}
 \eea

\section{Muon neutralino smuon and muon chargino sneutrino vertices.}

The interaction terms contain the lepton neutralino slepton and lepton chargino sneutrino vertices
are given as follows:
\bea L_{l\widetilde{l}\widetilde{V}} &=&-\frac{ig}{\sqrt{2}} \left( \overline{ L}
\lambda^a \widetilde{L} \overline{\lambda}_V^a -\overline{\widetilde{L}}\lambda^a L \lambda_V^a
\right) - ig^\prime \sqrt{\frac{1}{3}}\left[-\frac{1}{3}(\overline{L}\widetilde{L}\overline{\lambda}_B-\overline{\widetilde{L}}L\lambda_B)
+(\overline{l}^c \widetilde{l}^c \overline{\lambda}_B - \overline{\widetilde{l}}^c l^c \lambda_B)
\right],
 \label{mu1}\crn
 \\
  L_{l\widetilde{l}\widetilde{H}}&=&-\frac{\lambda_{1ab}}{3}\left(L_a
\widetilde{\rho}^\prime \widetilde{l}^c_b+\widetilde{L_a}\widetilde{\rho}^\prime l^c_b
\right)-\frac{\lambda_{3ab}}{3}\left(L_a \widetilde{\rho}
\widetilde{L_b}+\widetilde{L_a}\widetilde{\rho} L_b \right)
+ H.c. \label{mu2}
\eea

We would like to remind that the lepton number is conserved in the lepton sector at the tree level
and $\lambda_{3ab}$ are antisymmetric with $a$ and $b$ leading to the vanish of couplings $\lambda_{1ab},
\lambda_{3ab}$ if $a$ equals to $b$. We use
the notations $Y_\mu = \frac{\lambda_{122}}{3 }$  as given in \cite{LFV2}. Expanding the Eqs. (\ref{mu1}) and (\ref{mu2}), we rewrite only the muon-neutralino-smuon and muon-chargino-sneutrino interaction terms. All
relevant terms are given as
 \bea
  L_{\mu\widetilde{\mu}\widetilde{V}} &=&  \overline{\mu}_L
(\frac{ig^\prime}{3\sqrt{3}} \overline{\lambda}_B
+\frac{ig}{\sqrt{2}}(\overline{\lambda_A^3}-\frac{1}{\sqrt{3}}\overline{\lambda_A^8}))\widetilde{\mu}_L 
- \mu_L(\frac{ig^\prime}{3\sqrt{3}}\lambda_B+\frac{ig}{\sqrt{2}}(\lambda_A^3-\frac{1}{\sqrt{3}}\lambda_A^8))\widetilde{\mu}_L^* \crn
&-& i\frac{g^\prime }{\sqrt{3}}(\bar{\mu^c}\widetilde{\mu^c}\bar{\lambda_B}-\overline{\widetilde{\mu^c}} \mu^c \lambda_B)- ig \left( \overline{\mu}_L \overline{\widetilde{W}^+}\widetilde{ \nu}_{\mu L}- \mu_L\widetilde{W}^+ \widetilde{\nu}_{\mu L}^*\right) \crn
&+& ig\left( \overline{\mu}_L \overline{\widetilde{Y}^+}\widetilde{ \nu}_{\mu L}- \mu_L\widetilde{Y}^+
\widetilde{\nu}_{\mu L}^*\right).
\label{mu3}\\
L_{\mu\widetilde{\mu}\widetilde{H}}&=& -Y_\mu \{\mu_L \widetilde{\mu}_L^c
\widetilde{\rho}^{\prime o}+\widetilde{\mu}_L \widetilde{\rho}^{\prime o} \mu_L^c
+\widetilde{\nu}_{\mu L}\widetilde{\rho}_1^{\prime -}\mu_L^c +\widetilde{\nu}_{\mu L}^c
\widetilde{\rho}_2^{\prime -} \mu_{L}^c\}
+H.c. \label{mu4}
\eea
The Eqs. (\ref{mu3}),
(\ref{mu4}) can be written in the physical states as follows:
\bea
L_{\mu \widetilde{\mu} \chi_i}&=& \sum_{iA}\overline{\mu}(P_R N_{\chi_i \widetilde{\mu}_ A}^R
+P_L N_{\chi_i \widetilde{\mu}_A}^L)\widetilde{\mu}_A\overline{\chi}_i + \sum_{jB}\overline{\mu}(P_R
C_{\chi_j \widetilde{\nu}_B}^R + P_L C_{\chi_j \widetilde{\nu}_B}^L)\overline{\chi_j^+}
\widetilde{\nu}_A + H. c, \label{mug}
\eea
 where
  \bea N_{\chi_i \widetilde{\mu}_ A}^L &=&
\left(\frac{ig^\prime}{3 \sqrt{3}}N^{*}_{7i}+\frac{ig}{\sqrt{2}}N^{*}_{8i}-\frac{ig}{\sqrt{6}}N^{*}_{9i}\right)
\times (U_{\widetilde{\mu}})_{LA}-Y_\mu N^{*}_{6i}(U_{\widetilde{\mu}})_{RA}, \crn
 N_{\chi_i \widetilde{\mu}_ A}^R&=& -\frac{ig^\prime}{\sqrt{3}}N^*_{7i}(U_{\widetilde{\mu}})_{RA} - Y_\mu N^*_{6
i} (U_{\widetilde{\mu}})_{LA}, \crn
C_{\chi_j \widetilde{\nu}_B}^L &=& -ig\left ( V_{1j}(U_{\nu})_{LB}+
V_{2j}(U_{\nu})_{RB} \right), \crn
 C_{\chi_j \widetilde{\nu}_B}^R &=& -Y_\mu
(U_{3j}(U_{\nu})_{LA}+U_{4j}(U_{\nu})_{RA}).
\nn\eea

\section{The  SUSY contribution to the muon MDM}

The amplitude for the photon-muon-muon coupling in the  zero
limit of the momentum of photon can be written as:
\[
M_{fi}=ie\overline{u}\left[\gamma^\alpha
+ a_\mu\frac{i\sigma^{\alpha\beta}q_\beta}{2m_\mu} \right]uA_\alpha
\]

The magnetic dipole moment $a_\mu$ can be calculated
in both mass eigenstate and weak eigenstate. However in the next
section the magnetic dipole moment will be evaluated in the weak eigenstate,
 since in this basis the dependence of $a_\mu$ on SUSY parameters is more reveal
 than in mass eigenstate basis.

The SM prediction of the muon anomalous magnetic moment has been given in \cite{g2SM}
\[
a_\mu^{SM}=116591803(1)(42)(26)\times 10^{-11}
\]

The recent E821 experiments \cite{g2E821} have measured and take
into account correlations between systematic errors one finds
\bea
a_\mu^{E821}&=&116592091(54)(33)\times 10^{-11} \crn
&=&(116592091 \pm 63.3) \times 10^{-11}
\nn \eea
Hence
\bea
\Delta a_\mu(E821-SM)&=&288(63)(49)\times10^{-11} \crn
&=&(288 \pm 80)\times10^{-11}
\nn \eea

Then we have $\Delta a_\mu$ are $3.68 \times 10^{-9}$ and $2.08 \times 10^{-9}$  with the
 differences $3.6 \sigma$ and $2.0 \sigma$, respectively.

\subsection{Mass eigenstate}
The effect of supersymmetry on $a_\mu$ includes loops with charginos
 and neutralinos. The one-loop contributions to $a_\mu$ \cite{MSSMg-2},
 including the effects of possible complex phase, are:

\bea
\Delta a_\mu^{\chi^0\tilde{\mu}} &=&
m_\mu \sum_{i A} \Big\{-m_\mu
(|N^L_{\chi_i \tilde{\mu}_A}|^2
+ |N^R_{\chi_i \tilde{\mu}_A}|^2) m_{\tilde{\mu}A}^2
 \times J_5(m_{\chi_i}^2,m_{\tilde{\mu}A}^2,m_{\tilde{\mu}A}^2,
m_{\tilde{\mu}A}^2,m_{\tilde{\mu}A}^2) \crn
&+& m_{\chi_i} Re(N^{L*}_{\chi_i \tilde{\mu}_A}
N^R_{\chi_i \tilde{\mu}_A})
J_4(m_{\chi_i}^2,m_{\chi_i}^2,m_{\tilde{\mu}A}^2,m_{\tilde{\mu}A}^2) \Big\}
\crn
&=&\frac{1}{16\pi^2} m_\mu \sum_{iA}
\Bigg\{-\frac{m_\mu}{6m_{\tilde{\mu}A}^2(1-x_{\chi_i A})^4}
(|N^L_{\chi_i \tilde{\mu}_A}|^2 + |N^R_{\chi_i \tilde{\mu}_A}|^2)
 \times  (1-6x_{\chi_i A}+3x_{\chi_i A}^2 \crn
 &+& 2x_{\chi_i A}^3-6x_{\chi_i A}^2\ln x_{\chi_i A})
- \frac{m_{\chi_i}}{m_{\tilde{\mu}A}^2(1-x_{\chi_i A})^3}Re(N^{L*}_{\chi_i \tilde{\mu}_A} N^R_{\chi_i \tilde{\mu}_A})  \crn &\times & (1-x_{\chi_i A}^2+2x_{\chi_i A}\ln x_{\chi_i A})
\Bigg\},\crn
\label{g-2_nAX}
\Delta a_\mu^{\chi^\pm\tilde{\nu}} &=&
m_\mu \sum_{i}\Big[ m_\mu
(|C^L_{\chi_i \tilde{\nu}}|^2 + |C^R_{\chi_i \tilde{\nu}}|^2)\times
\{  m_{\tilde{\nu}}^2 J_5(m_{\chi_i^\pm}^2,m_{\chi_i^\pm}^2,m_{\chi_i^\pm}^2
,m_{\chi_i^\pm}^2,m_{\tilde{\nu}}^2)\crn
&+&J_4(m_{\chi_i^\pm}^2,m_{\chi_i^\pm}^2,m_{\chi_i^\pm}^2,m_{\chi_i^\pm}^2),
-J_4(m_{\chi^\pm X}^2,m_{\chi^\pm X}^2,m_{\chi^\pm X}^2
m_{\tilde{\nu}}^2) \} \crn
&-&2m_X Re(C^{L*}_{\chi_i \tilde{\nu}} C^R_{\chi_i \tilde{\nu}})
J_4 (m_{\chi_i^\pm X}^2,m_{\chi_i^\pm X}^2,m_{\chi_i^\pm X}^2
,m_{\tilde{\nu}}^2) \Big]
\crn
 &=& \frac{1}{16\pi^2} m_\mu \sum_{i}
\Bigg\{\frac{m_\mu}{3m_{\tilde{\nu}}^2(1-x_{\chi_i^\pm})^4}
(|C^L_{\chi_i \tilde{\nu}}|^2 + |C^R_{\chi_i \tilde{\nu}}|^2)\crn
&\times& \left(
1 + \frac{3}{2} x_{\chi_i^\pm} - 3x_{\chi_i^\pm}^2
+ \frac{1}{2}x_{\chi_i^\pm}^3
+ 3x_{\chi_i^\pm}\ln x_{\chi_i^\pm} \right) \crn
&-&\frac{3m_{\chi^\pm }}{m_{\tilde{\nu}}^2
(1-x_{\chi_i^\pm})^3} Re(C^{L*}_{\chi_i \tilde{\nu}} C^R_{\chi_i \tilde{\nu}})
\left( 1 - \frac{4}{3}x_{\chi_i^\pm} + \frac{1}{3}x_{\chi_i^\pm}^2
+ \frac{2}{3}\ln x_{\chi_i^\pm} \right)
\Bigg\},
\label{g-2_cAX}
\eea

where $x_{\chi_i^\pm}=m^2_{\chi_i^\pm}/m_{\tilde{\nu}}^2, \,
 x_{\chi_i A} =m^2_{\chi_i}/m^2_{\tilde{\mu_A}}$.

 Based on the contributions of SUSYE331 to the muon MDM
 given in the Eqs. (\ref{g-2_nAX}), (\ref{g-2_cAX}), it is hard to
 see the effects of the SUSYE331 parameter space to the muon MDM,  in
 particular,  the role of $\tan \gamma $. To assess the effects of
 SUSYE331 parameter space to the muon MDM, it is convenient to use
 the mass insertion method to calculate the diagrams. In next
 part, let us consider the muon MDM based on the weak eigenstate.

\subsection{Weak eigenstate}

\begin{figure}[!ht]
    \centering
    \includegraphics[width=0.5\textwidth]{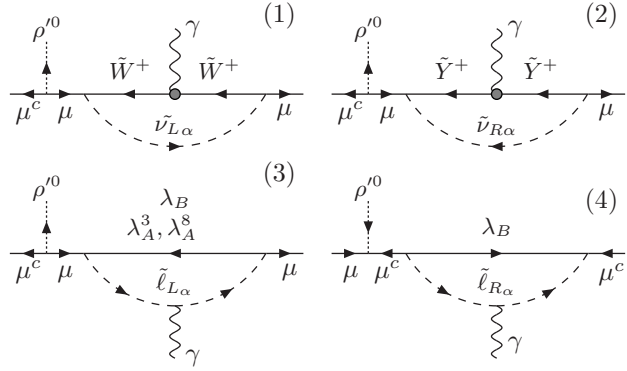}
    \caption{Contribution to $a_{\mu L}^{(a)}[1-3]$ and $a_{\mu R}^{(a)}[4]$}
    \label{Amua-diagrams}
\end{figure}

\begin{figure}[!ht]
    \centering
    \includegraphics[width=0.5\textwidth]{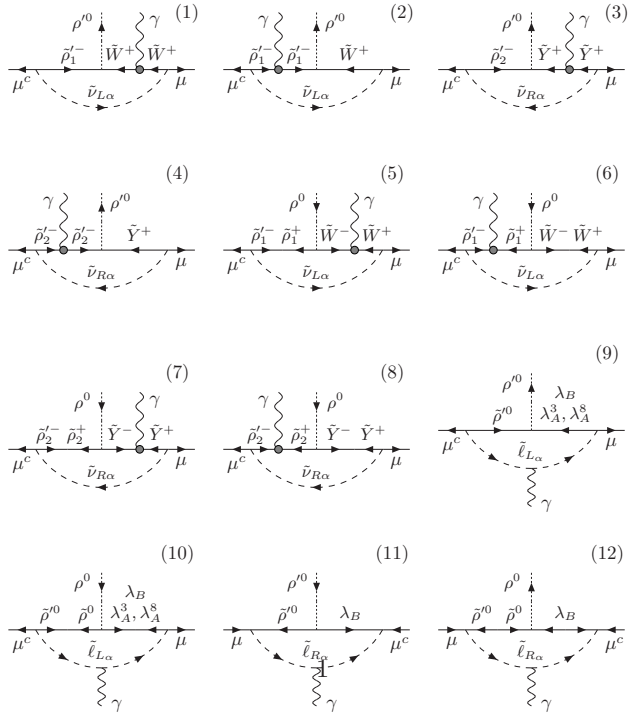}
    \caption{Contribution to $a_{\mu L}^{(b)}[1-10]$ and $a_{\mu R}^{\mu(b)}[11-12]$}
    \label{Amub-diagrams}
\end{figure}

\begin{figure}[!ht]
    \centering
    \includegraphics[width=0.5\textwidth]{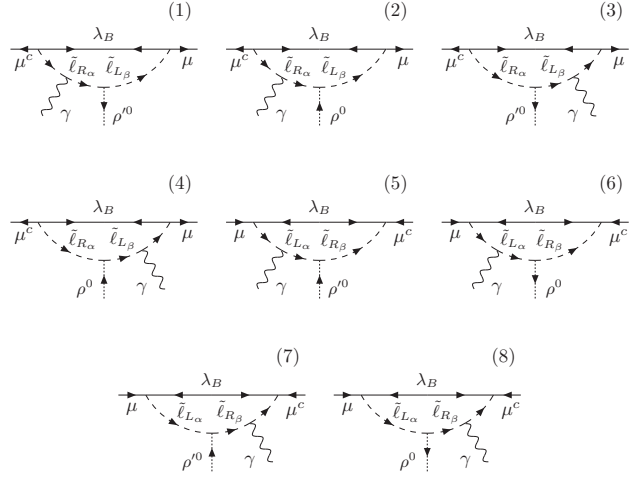}
    \caption{Contribution to $a_{\mu L}^{(c)}[1-6]$ and $a_{\mu R}^{(c)}[7-8]$}
    \label{Amuc-diagrams}
\end{figure}

In this section, let us consider the SUSY contribution to the muon MDM  by using the mass insertion
method to calculate the diagrams in Fig. \ref{Amua-diagrams},\ref{Amub-diagrams},\ref{Amuc-diagrams}.
 The contributions to  $a_\mu$ can be separated into six parts:
 $ a_\mu^{SUSYE331}=a^{(a)}_{\mu L}+a^{(a)}_{\mu R}+a^{(b)}_{\mu L}+a^{(b)}_{\mu R}+a^{(c)}_{\mu
 L}+a^{(c)}_{\mu R}$,
 where diagrams involving each part are expressed in
  three Figs.\ref{Amua-diagrams}, \ref{Amub-diagrams} and
 \ref{Amuc-diagrams}. Their contributions are given as follows:
\bea
 a^{(a)}_{\mu L}&=&-\frac{g^2 m_\mu^2}{8\pi^2}\left[\frac{m^2_{\tilde{l}_{L2}}c_L^2}{3}
 J_5(m^2_{\lambda},m^2_{\tilde{l}_{L2}},m^2_{\tilde{l}_{L2}},
 m^2_{\tilde{l}_{L2}},m^2_{\tilde{l}_{L2}} )  \right. \crn
 &-& \left. \frac{m^2_{\lambda}c^2_{\nu_L}}{2} ~J_5(m^2_{\lambda},m^2_{\lambda},m^2_{\lambda},
 m^2_{\lambda},m^2_{\tilde{\nu}_{L2}})\right]\crn
 &+&\frac{g^2c_{\nu_R}^2 m_\mu^2}{8\pi^2}~\frac{m^2_{\lambda}}{2}
  ~J_5(m^2_{\lambda},m^2_{\lambda},m^2_{\lambda},
 m^2_{\lambda},m^2_{\tilde{\nu}_{R2}}) \crn
 &-& \frac{g^{\prime2}c_L^2 m^2_\mu}{8\pi^2}~\frac{m^2_{\tilde{l}_{L2}}}{54}
 J_5(m^2_B,m^2_{\tilde{l}_{L2}},m^2_{\tilde{l}_{L2}},
 m^2_{\tilde{l}_{L2}},m^2_{\tilde{l}_{L2}} )\crn
 &+& \left (L_2\rightarrow
 L_3,R_2\rightarrow R_3, c_L^2 \rightarrow s_L^2, c_{\nu_R}^2\rightarrow s_{\nu_R}^2\right),
 \label{amuaL}\\
 a^{(a)}_{\mu R}&=&-\frac{g^{\prime2}c_R^2 m_\mu^2}{8\pi^2}\frac{m^2_{\tilde{l}_{R2}}}{6}
 J_5(m^2_B,m^2_{\tilde{l}_{R2}},m^2_{\tilde{l}_{R2}},
 m^2_{\tilde{l}_{R2}},m^2_{\tilde{l}_{R2}} )+\left( R_2\rightarrow R_3,c_R^2 \rightarrow s_R^2 \right),
 \label{amuaR} \\
a^{(b)}_{\mu L} &=&
\frac{g^{2}c_{\nu_L}^2m_\mu^2}{8 \pi^2} m^4_{\tilde{\nu}_{L2}}I_5(m_{\lambda}^2,\mu^2_{\rho},
m^2_{\tilde{\nu}_{L2}},m^2_{\tilde{\nu}_{L2}},m^2_{\tilde{\nu}_{L2}})  \crn
&+& \frac{g^{2}c_{\nu_R}^2 m_\mu^2}{8 \pi^2}
m^4_{\tilde{\nu}_{R2}}I_5(m_{\lambda}^2,\mu^2_{\rho},
m^2_{\tilde{\nu}_{R2}},m^2_{\tilde{\nu}_{R2}},m^2_{\tilde{\nu}_{R2}})
\crn
&-& \frac{g^2c_{\nu_{L}}^2 m_\mu^2}{8\pi^2} m_{\lambda}~\mu_\rho~\tan\gamma
\left[ J_5(m_{\lambda}^2,m_{\lambda}^2,\mu^2_{\rho},\mu^2_{\rho},m^2_{\tilde{\nu}_{L2}})
 +J_5(m_{\lambda}^2,m_{\lambda}^2,m_{\lambda}^2,\mu^2_{\rho},m^2_{\tilde{\nu}_{L2}}) \right. \crn
&+& \left. J_5(m_{\lambda}^2,\mu^2_{\rho},\mu^2_{\rho},\mu^2_{\rho},m^2_{\tilde{\nu}_{L2}}) \right]
- \frac{g^2c_{\nu_{R}}^2 m_\mu^2}{8\pi^2} m_{\lambda}~\mu_\rho~\tan\gamma
\left[ J_5(m_{\lambda}^2,m_{\lambda}^2,\mu^2_{\rho},\mu^2_{\rho},m^2_{\tilde{\nu}_{R2}})\right. \crn
&+&\left. J_5(m_{\lambda}^2,m_{\lambda}^2,m_{\lambda}^2,\mu^2_{\rho},m^2_{\tilde{\nu}_{R2}})
+J_5(m_{\lambda}^2,\mu^2_{\rho},\mu^2_{\rho},\mu^2_{\rho},m^2_{\tilde{\nu}_{R2}}) \right]
\crn
&+& \frac{g^{2}c_L^2 m_\mu^2}{8 \pi^2}~m^2_{\tilde{l}_{L2}}\frac{2}{3} \left[
J_5(m_{\lambda}^2,\mu^2_{\rho},
m^2_{\tilde{l}_{L2}},m^2_{\tilde{l}_{L2}},m^2_{\tilde{l}_{L2}})  \right. \crn
&-& \left.   m_{\lambda}~\mu_{\rho}\tan\gamma~I_5(m_{\lambda}^2,\mu^2_{\rho},
m^2_{\tilde{l}_{L2}},m^2_{\tilde{l}_{L2}},m^2_{\tilde{l}_{L2}})\right]
 \crn
  &-& \frac{g^{\prime2}c_L^2 2 m_\mu^2}{8 \pi^2}~m^2_{\tilde{l}_{L2}}
\frac{2}{27}
\left[J_5(m_B^2,\mu^2_{\rho},m^2_{\tilde{l}_{L2}},m^2_{\tilde{l}_{L2}},m^2_{\tilde{l}_{L2}}
)  \right. \crn &-& \left.   m_B~\mu_\rho~\tan\gamma~
 I_5(m_B^2,\mu^2_{\rho},m^2_{\tilde{l}_{L2}},m^2_{\tilde{l}_{L2}},m^2_{\tilde{l}_{L2}})
  \right]\crn
  &+&[L_2\rightarrow L_3,R_2\rightarrow R_3, c^2_L \rightarrow s^2_L, c^2_{\nu_R} \rightarrow
   s^2_{\nu_R}, c^2_{\nu_L} \rightarrow s^2_{\nu_L}],
   \label{amubL}\\
 a^{(b)}_{\mu R}&=&-\frac{g^{\prime2}c_R^2 m_\mu^2}{8
\pi^2} m^2_{\tilde{l}_{R2}}\frac{2}{9}
\left[-J_5(m_B^2,\mu^2_{\rho},m^2_{\tilde{l}_{R2}},m^2_{\tilde{l}_{R2}},m^2_{\tilde{l}_{R2}}
)  \right. \crn &+& \left.  m_B~\mu_{\rho}\tan\gamma~
I_5(m_B^2,\mu^2_{\rho},m^2_{\tilde{l}_{R2}},m^2_{\tilde{l}_{R2}},m^2_{\tilde{l}_{R2}} )
\right] \crn
&+& [R_2\rightarrow R_3, c^2_R \rightarrow s^2_R].
 \label{amubR}\\
a^{(c)}_{\mu LR}&=&\frac{g^{\prime 2}m_\mu^2}{8 \pi^2} \frac{m_B^3}{9}
\left[ \left( \frac{m_\tau}{m_\mu} \left[ c_R^2 s_L c_L A^R_{\mu \tau}
 \right. \right. \right. \left. \left. \left. + s_R c_R c_L^2 A^L_{\mu \tau}
 +s_R c_Rs_Lc_L \left( A_\tau +\frac{\mu_\rho \tan \gamma}{2}\right)\right] \right. \right. \crn
 &+& \left. \left.  c^2_Rc^2_L  \left[A_\mu+\frac{\mu_\rho \tan \gamma}{2}\right] \right)
 I_5(m_B^2,m_B^2,m_B^2,m^2_{\tilde{l}_{L2}},
 m^2_{\tilde{l}_{R2}}) \right.\crn
& + & \left. \left( \frac{m_\tau}{m_\mu} \left( -c_R^2 s_L c_L A^R_{\mu \tau}+
 s_Rc_Rs^2_L A^L_{\mu \tau} -s_Rc_R s_L c_L \left[ A_\tau
 +\frac{\mu_\rho \tan \gamma}{2} \right]\right) \right. \right. \crn
  &+& \left. \left. c^2_Rs^2_L\left[A_\mu+\frac{\mu_\rho \tan \gamma}{2}\right]\right)
   I_5 (m_B^2,m_B^2,m_B^2,m^2_{\tilde{l}_{L3}},
 m^2_{\tilde{l}_{R2}})\right.\crn
 & + & \left. \left( \frac{m_\tau}{m_\mu}
 \left( s_R^2s_Lc_L A^R_{\mu \tau} -s_Rc_Rc_L^2A^L_{\mu \tau}-s_Rc_Rs_Lc_L
 \left[ A_\tau +\frac{\mu_\rho \tan \gamma}{2}\right]\right) \right. \right. \crn
 &+&  \left. \left. s_R^2c_L^2
 \left[A_\mu +\frac{\mu_\rho \tan \gamma}{2} \right]\right)I_5
 (m_B^2,m_B^2,m_B^2,m^2_{\tilde{l}_{L2}},
 m^2_{\tilde{l}_{R3}})\right. \crn
 &+&  \left. \left(\frac{m_\tau}{m_\mu}
 \left[-s_R^2s_Lc_LA^R_{\mu \tau}-s_Rc_Rs_L^2 A^L_{\mu \tau}
 +s_Rc_Rs_Lc_L \left( A_\tau +\frac{\mu_\rho \tan \gamma}{2}\right)\right]  \right. \right. \crn
 &+& \left. \left. s_R^2s_L^2
  \left[
 A_\mu +\frac{\mu_\rho \tan \gamma}{2} \right]\right)I_5(m_B^2,m_B^2,m_B^2,
 m^2_{\tilde{l}_{L3}},
 m^2_{\tilde{l}_{R3}})\right],
  \label{amuc}
\eea where $s_L, c_L$ and $s_R, c_R$ are the mixing angles that related between
flavour states $\tilde{\mu}_L, \tilde{\tau}_L,\tilde{\mu}_L^c, \tilde{\tau}_L^c$
and the mass states
 $\tilde{l}_{L_2},\tilde{l}_{L_3},\tilde{l}_{R_2},\tilde{l}_{R_3}$, namely
\bea
 \tilde{\mu} _L&=&c_L \tilde{l}_{L_2} -s_L \tilde{l}_{L_3},\hs
 \tilde{\tau}_L= s_L \tilde{l}_{L_2}+c_L \tilde{l}_{L_3}, \crn
  \tilde{\mu} _L^c&=&c_R \tilde{l}_{R_2} -s_R \tilde{l}_{R_3}, \hs
  \tilde{\tau}_L^c=s_R \tilde{l}_{R_2}+c_R \tilde{l}_{R_3}.
\nn \eea
with $c_L= \cos \theta_L, s_L= \sin \theta_L, c_R=\cos \theta_R, s_R=\sin \theta_R$.
$A_\mu, A_{\mu \tau}^{L,R}$ are the couplings of the smuon-smuon-neutral Higgs,
smuon-stauon-neutral Higgs, respectively. More details on the symbol, an interested reader can see in
 \cite{LFV1}.

\section{Numerical Calculation}

The full parameter space of the SUSYE331 model contains dozens of parameters
however we can classify in categories: $B/ \mu$-term: $\mu_\rho$;
the ratio of two vacua: $\tan{\gamma}$; gauginos mass $m_B$, $m_\lambda$;
right-handed slepton mass: $m_{\tilde{l}_{R_2}}$, $m_{\tilde{\nu}_{R_2}}$,
$m_{\tilde{l}_{R_3}}$,$m_{\tilde{\nu}_{R_3}}$ ; left-handed slepton mass:
$m_{\tilde{l}_{L_2}}$, $m_{\tilde{\nu}_{L_2}}$,
$m_{\tilde{l}_{L_3}}$,$m_{\tilde{\nu}_{L_3}}$ ;
mixing terms: $A_\mu$, $A_\tau$, $A^L_{\tau \mu}$, $A^R_{\tau \mu}$.
To simplify our calculation we  can first make a rough estimation by taking
the limit of $m_{\tilde{L}}$, $m_B$, $m_\lambda$, $\mu_\rho$ to $m_{SUSY}$
and since  $A_\tau$, and $A^{L,R}_{\mu \tau}$ are non-diagonal terms in
mixing matrix meaning $A_\tau$, and $A^{L,R}_{\mu \tau}$ are very small
then we can approximate $A_\tau=A^{L,R}_{\mu \tau}=0$.

In this limit, the analytical expressions
(\ref{amuaL}),(\ref{amuaR}),(\ref{amubL}),(\ref{amubR}),(\ref{amuc})
can be written simply as follows
\bea
a^{(a)}_{\mu L}&=& -\frac{1}{18} \frac{g^2}{8 \pi^2}\frac{m_\mu^2}{m^2_{SUSY}}
+\frac{1}{54}\frac{1}{12} \frac{g'^2}{8 \pi^2}\frac{m_\mu^2}{m^2_{SUSY}}, \crn
a^{(a)}_{\mu R}&=& \frac{1}{6}\frac{1}{12} \frac{g'^2}{8 \pi^2}\frac{m_\mu^2}
{m^2_{SUSY}}, \crn
a^{(b)}_{\mu L}&=& \frac{1}{12}\frac{g^2}{4 \pi^2}\frac{m_\mu^2}{m^2_{SUSY}}
 + \frac{1}{4}\frac{g^2}{4 \pi^2}\frac{m_\mu^2}{m^2_{SUSY}}
 sign(\mu_\rho)\tan{\gamma} -  \frac{1}{12} \frac{2}{3}\frac{g^2}{8 \pi^2}\frac{m_\mu^2}{m^2_{SUSY}}
(1+sign(\mu_\rho)\tan{\gamma})
\crn
&+&\frac{1}{12} \frac{4}{27}\frac{g'^2}{8 \pi^2}\frac{m_\mu^2}{m^2_{SUSY}}
(1+sign(\mu_\rho)\tan{\gamma}), \nonumber \\
a^{(b)}_{\mu R}&=& -\frac{1}{12} \frac{2}{9}\frac{g'^2}{8 \pi^2}
\frac{m_\mu^2}{m^2_{SUSY}}(1+sign(\mu_\rho)\tan{\gamma}),\crn
a^{(c)}_{\mu LR}&=&\frac{1}{12}\frac{1}{9}\frac{g'^2}
{8 \pi^2}\frac{m_\mu^2}{m^3_{SUSY}}Re\left[
A_\mu+ \frac{m_{SUSY}sign(\mu_\rho)\tan{\gamma}}{2} \right].
\nn
\eea
If we assume $A_\mu=0$ then summing all above terms we have
\bea
\Delta a_\mu^{total} &=&-\frac{1}{36}\frac{g^2}{8 \pi^2}\frac{m^2_\mu}{m^2_{SUSY}}
 + \frac{1}{108}\frac{g'^2}{8 \pi^2}\frac{m^2_\mu}{m^2_{SUSY}} 
+ \frac{4}{9}\frac{g^2}{8\pi^2}sign(\mu_\rho)\tan{\gamma}\frac{m_\mu^2}{m^2_{SUSY}} \crn
&+& \frac{1}{648}\frac{g'^2}{8\pi^2}sign(\mu_\rho)\tan{\gamma}
\frac{m_\mu^2}{m^2_{SUSY}}.
\label{amu-analytic}
\eea

\begin{figure*}[!ht]
    \centering
    \includegraphics[width=0.7\textwidth]{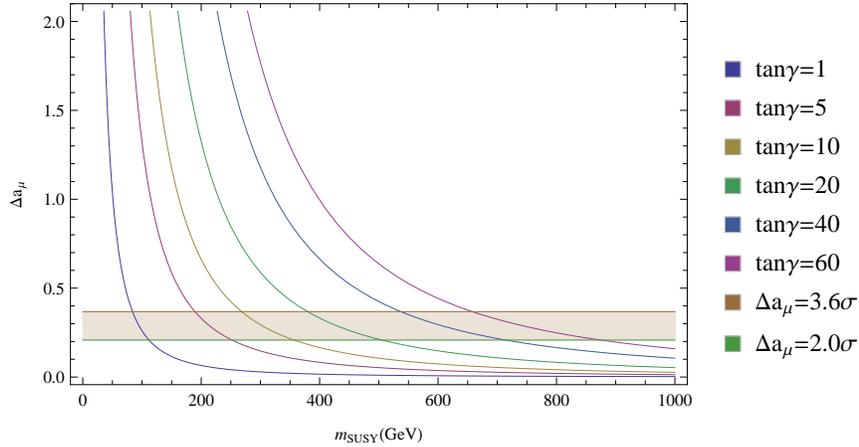}
    \caption{$\Delta a_\mu$ plot against $M_{SUSY}$}
    \label{AmutotalPlot}
\end{figure*}
  First let us numerically estimate $\Delta a_\mu$ by using the
  Eq.(\ref{amu-analytic}). From Eq.  (\ref{amu-analytic}) we can conclude that $\Delta a_\mu$ is
 the same sign with $sign(\mu_\rho)$. In Fig. \ref{AmutotalPlot} we plot
 the discrepancy of the  muon MDM between experimental  data and that predicted by
  the SM. For our convenience we have scale up the value of MDM by factor of $10^8$
  through out this paper. The shade region is  $1.6\, \sigma$ difference with
  upper bound and lower bound are $2.0-3.6\, \sigma$, respectively.
  The muon MDM is investigated with different values of
  $\tan{\gamma}=1,5,10,20,40,60 $. As we can see that to explain
  the $\geq 3.6 \sigma$ difference the mass of supersymmetric particle
  can be as small as $ \approx 75$ GeV provided a value of $\tan{\gamma}=1$.
  This is because in the SUSYE331 model, the number of new particles
  is increased comparing to the MSSM. Therefore the contribution to
  the total value of MDM is large even in the case of small $\tan{\gamma}$.
  When $\tan \gamma $ is large ($\tan \gamma$=60) the mass of the
  SUSY particle $m_{SUSY}$ is limited to 900 GeV in order to address
  the $2.0\,  \sigma$ discrepancy. As pointed out in \cite{Czarnecki} the simple
  extended of gauge symmetry model of the SM cannot address the problem of
  MDM because of the dampened term $\frac{m^2_\mu}{M^2_{NP}}$ where $M_{NP}$
  is the mass of the new physics particle. However in the supersymmetric
  version of the 3-3-1 model, the contribution to the MDM of the new particle
  is enhanced with a factor of the ratio of two vacuum $\tan \gamma$.
  Therefore the issue of the MDM can be addressed with an suitable
  value of $\tan \gamma$.

\begin{figure*}[!ht]\label{Amu-murho-mG-12}
\centering
\begin{minipage}{.5\textwidth}
  \centering
  \includegraphics[width=1.0\linewidth]{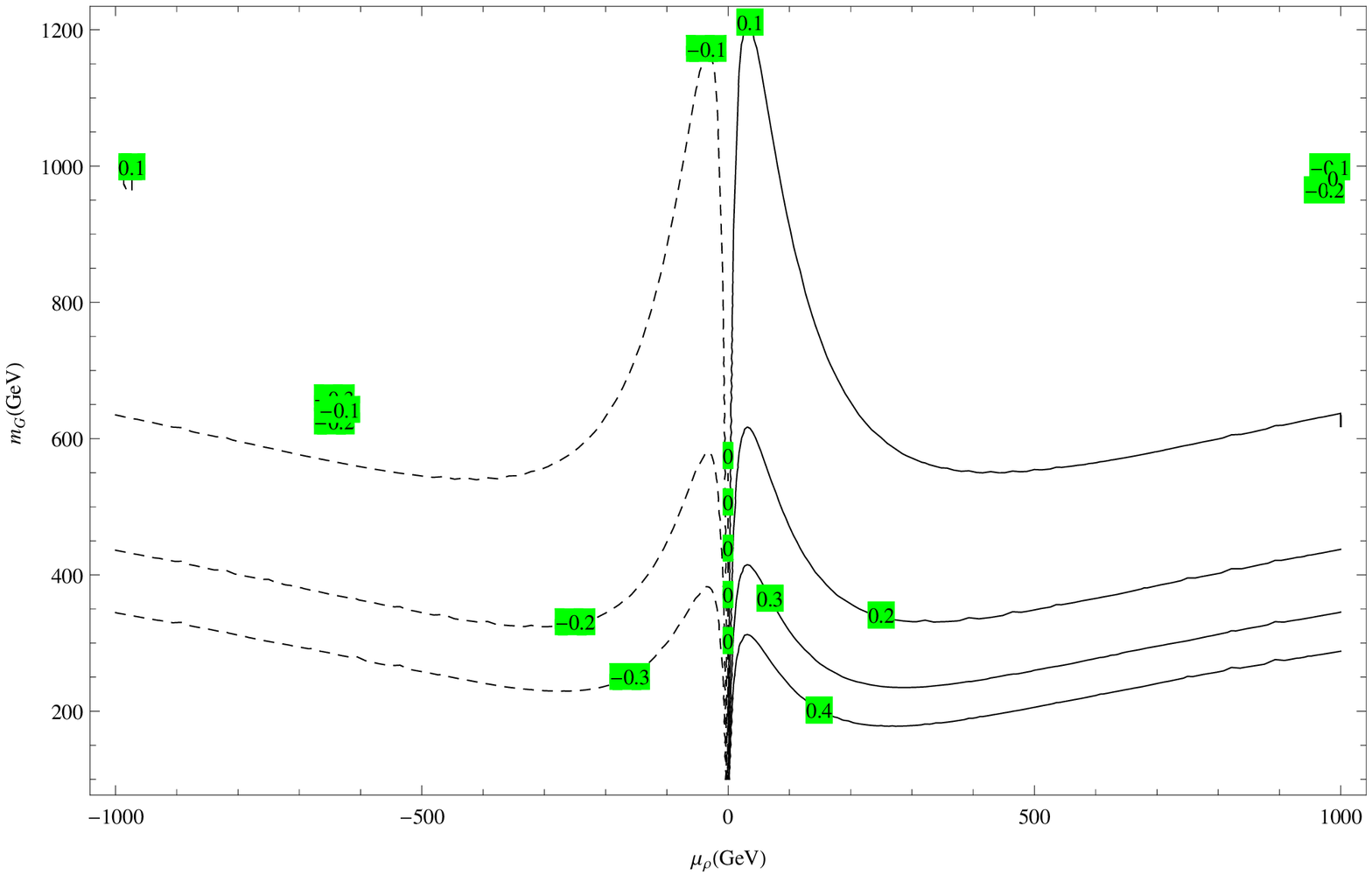}
    \caption{$\Delta a_\mu$ plotted against $\mu_\rho$ and $m_G$\\
$m_{\tilde{l}_2}=m_{\tilde{l}_{L_2}}=m_{\tilde{l}_{R_2}}=m_{\tilde{\nu}_{L_2}}=m_{\tilde{\nu}_{R_2}}$,\\
 $m_{\tilde{l}_3}=m_{\tilde{l}_{L_3}}=m_{\tilde{l}_{R_3}}=m_{\tilde{\nu}_{L_3}}=m_{\tilde{\nu}_{R_3}}$,\\
    $\tan \gamma=5$, $m_{\tilde{l}_2} = 100$ GeV, $m_{\tilde{l}_3} = 1$ TeV,\\
     $m_B=m_\lambda=m_G$,\\
  $\theta_L=\theta_R=\frac{\pi}{4}$,
 $\theta_{\nu_L}=\theta_{\nu_R}=\frac{\pi}{4} $.   }%
    \label{Amu-mu-mG-12-a}
\end{minipage}%
\begin{minipage}{.5\textwidth}
  \centering
  \includegraphics[width=1.0\linewidth]{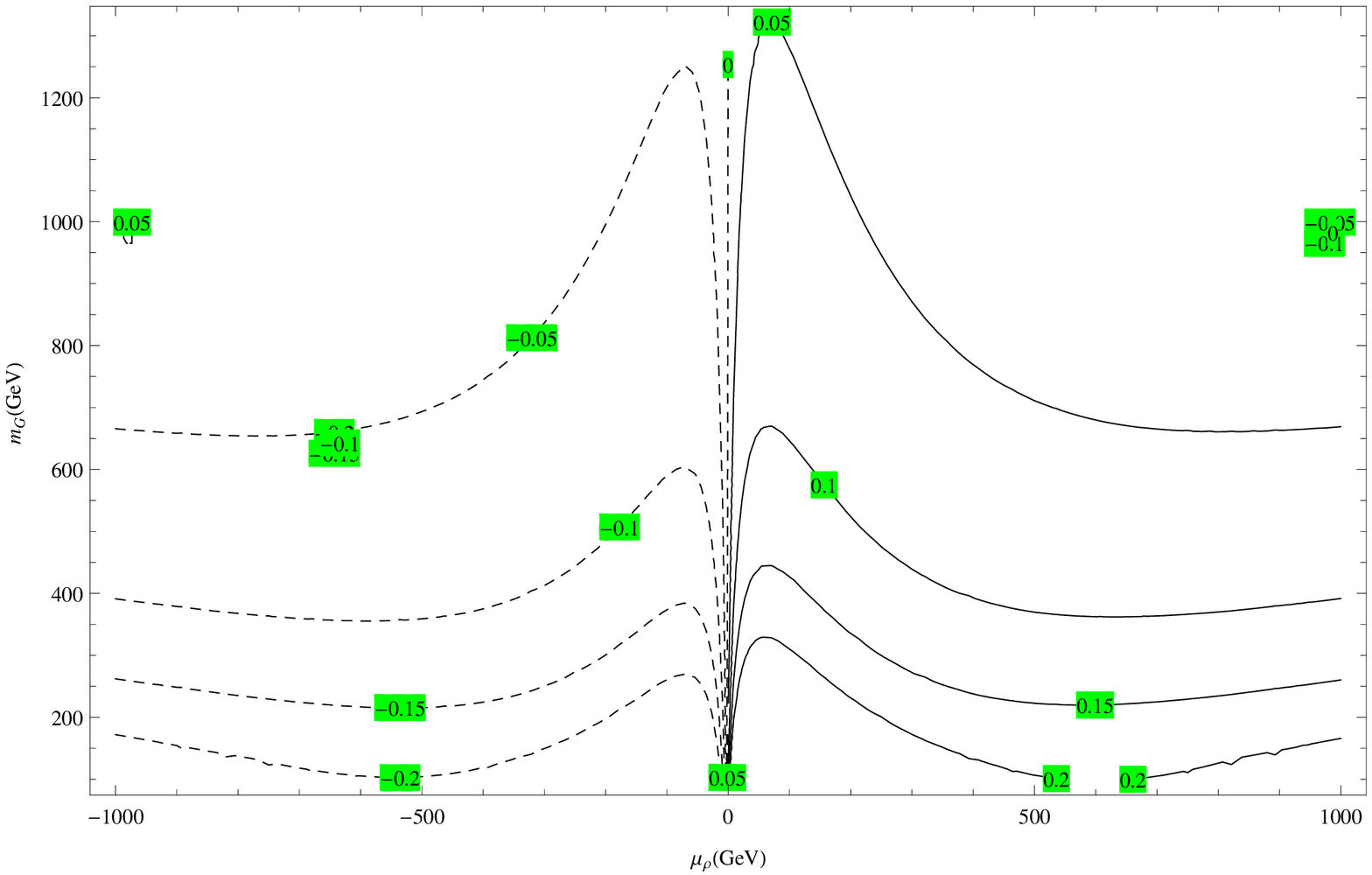}
  \caption{$\Delta a_\mu$ plotted against $\mu_\rho$ and $m_G$\\
$m_{\tilde{l}_2}=m_{\tilde{l}_{L_2}}=m_{\tilde{l}_{R_2}}=m_{\tilde{\nu}_{L_2}}=m_{\tilde{\nu}_{R_2}}$,\\
 $m_{\tilde{l}_3}=m_{\tilde{l}_{L_3}}=m_{\tilde{l}_{R_3}}=m_{\tilde{\nu}_{L_3}}=m_{\tilde{\nu}_{R_3}}$,\\
    $\tan \gamma=5$, $m_{\tilde{l}_2} = 200$ GeV, $m_{\tilde{l}_3} = 1$ TeV,\\
     $m_B=m_\lambda=m_G$,\\
  $\theta_L=\theta_R=\frac{\pi}{4}$,
 $\theta_{\nu_L}=\theta_{\nu_R}=\frac{\pi}{4} $.   }%
    \label{Amu-mu-mG-12-b}
    \end{minipage}
\end{figure*}

 Next, we show the SUSYE331 contribution to the muon MDM by using
 the analytical expressions
 (\ref{amuaL})-(\ref{amuc}) and fixing the value of $\tan \gamma$ and
 slepton masses.
We have assumed that
$m_{\tilde{l}_2}=m_{\tilde{l}_{L_2}}=m_{\tilde{l}_{R_2}}=m_{\tilde{\nu}_{L_2}}=m_{\tilde{\nu}_{R_2}}$  and
$m_{\tilde{l}_3}=m_{\tilde{l}_{L_3}}=m_{\tilde{l}_{R_3}}=m_{\tilde{\nu}_{L_3}}=m_{\tilde{\nu}_{R_3}}$.
Mass hierarchy between the second and the third generation is taken in to account.
Since the  MSSM is embedded in the SUSYE331 then we can
take the constraint on smuon mass \cite{PDG} where mass of smuon is
 greater than 91 GeV (ABBIENDI 04).
 We have approximated the mass of the second generation
  $m_{\tilde{l}_2} = 100$ GeV and other cases $m_{\tilde{l}_2} =200$ GeV
 while the mass of the third generation $m_{\tilde{l}_3}$ about 1TeV.
The results obtained in Fig. \ref{AmutotalPlot}  show that if the SUSY masses are fixed
in $100-200$ GeV, the values of $\tan \gamma$ equals 5 in order to fit the
experimental results.
Hence we study the SUSY contribution to the muon MDM for
fixing  values of slepton masses and in the $m_G$ and $\mu_\rho$ plane.
In Figs. \ref{Amu-mu-mG-12-a} and  \ref{Amu-mu-mG-12-b} we plot the results for
$\tan \gamma =5$, bino  and gauginos masses  are
assumed  to be equal gauginos mass, $m_B=m_\lambda=m_G$.
The mixing is assumed  maximal, $\theta_L=\theta_R=\frac{\pi}{4}$,
 $\theta_{\nu_L}=\theta_{\nu_R}=\frac{\pi}{4} $.
The results given in Fig. \ref{Amu-mu-mG-12-a} are plotted for
$m_{\tilde{l}_2}=100\,  {\rm GeV}, m_{\tilde{l}_3}=1$ TeV  while the results given in the  Fig.
 \ref{Amu-mu-mG-12-b} are plotted for $m_{\tilde{l}_2}=200\,  {\rm GeV},
 m_{\tilde{l}_3}=1$ TeV.
We plot for both negative and positive values  of $\mu_\rho$.
 There is a slightly asymmetry in the graph which caused by terms which
 do not depend on $\mu_\rho$. We  have imposed the condition of maximum
 value of $|\mu_\rho|\le 1500$ GeV to avoid fine-tuning requirement for
 the Higgs potential. From Fig. \ref{Amu-mu-mG-12-a} we can see that in
 order to address the anomalous of the  muon MDM data, the mass of the gauginos
  are in the range of $ 200 \le m_G \le 700$ GeV and
because of the set up of the masses, $\tan \gamma=5$ is
the minimum value to satisfy experimental  discrepancy $2-3.6\, \sigma$.
We can also learn that the value of MDM is inverse proportional to
the value of $\mu_\rho$. In the case when the mass of the
second generation is taken to be 200 GeV Fig. \ref{Amu-mu-mG-12-b},
the value of MDM merely reach  $2.0\,  \sigma$ anomaly of experimental data
which set the upper bound of the second generation mass to 200 GeV given
the mass of the third generation 1TeV and $\tan{\gamma}=5$. The results
given in  Figs.
\ref{Amu-mu-mG-12-a} and  \ref{Amu-mu-mG-12-b} show that if we take a
larger value
of $m_{\tilde{l_2}}$, the SUSYE331 contribution to the muon MDM is enhanced in the
small $m_G$ region.

\begin{figure*}[!ht]\label{Amu-murho-mG-34}
\centering
\begin{minipage}{.5\textwidth}
  \centering
  \includegraphics[width=1.0\linewidth]{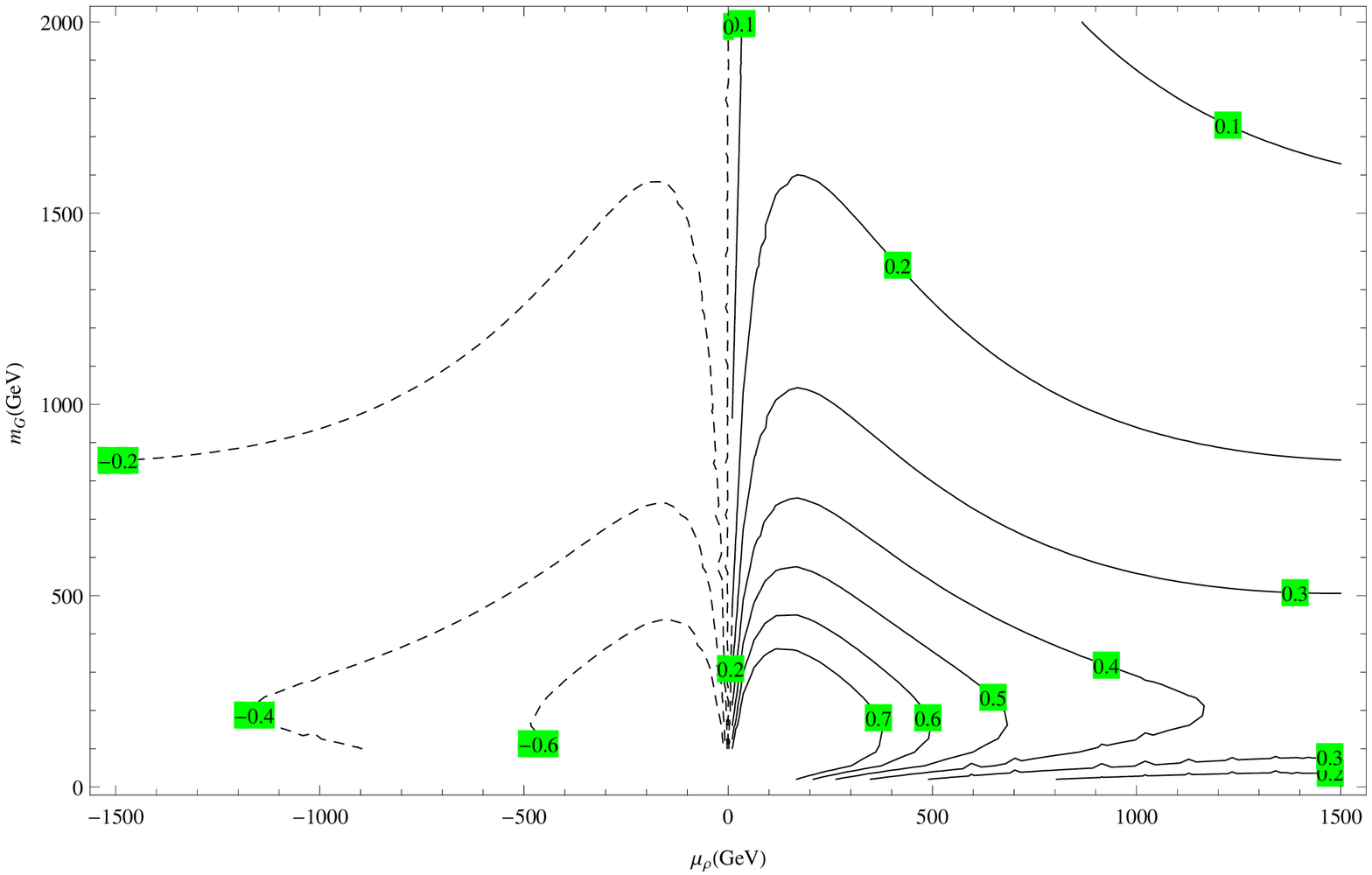}
\caption{$\Delta a_\mu$ plotted against $\mu_\rho$ and $m_G$\\
$m_{\tilde{l}_2}=m_{\tilde{l}_{L_2}}=m_{\tilde{l}_{R_2}}=m_{\tilde{\nu}_{L_2}}=m_{\tilde{\nu}_{R_2}}$,\\
 $m_{\tilde{l}_3}=m_{\tilde{l}_{L_3}}=m_{\tilde{l}_{R_3}}=m_{\tilde{\nu}_{L_3}}=m_{\tilde{\nu}_{R_3}}$,\\
    $\tan \gamma=60$, $m_{\tilde{l}_2} = 500$ GeV, $m_{\tilde{l}_3} = 2$ TeV,\\
     $m_B=m_\lambda=m_G$,\\
  $\theta_L=\theta_R=\frac{\pi}{4}$,
 $\theta_{\nu_L}=\theta_{\nu_R}=\frac{\pi}{4} $.   }%
    \label{Amu-mu-mG-34-a}
\end{minipage}%
\begin{minipage}{.5\textwidth}
  \centering
  \includegraphics[width=1.0\linewidth]{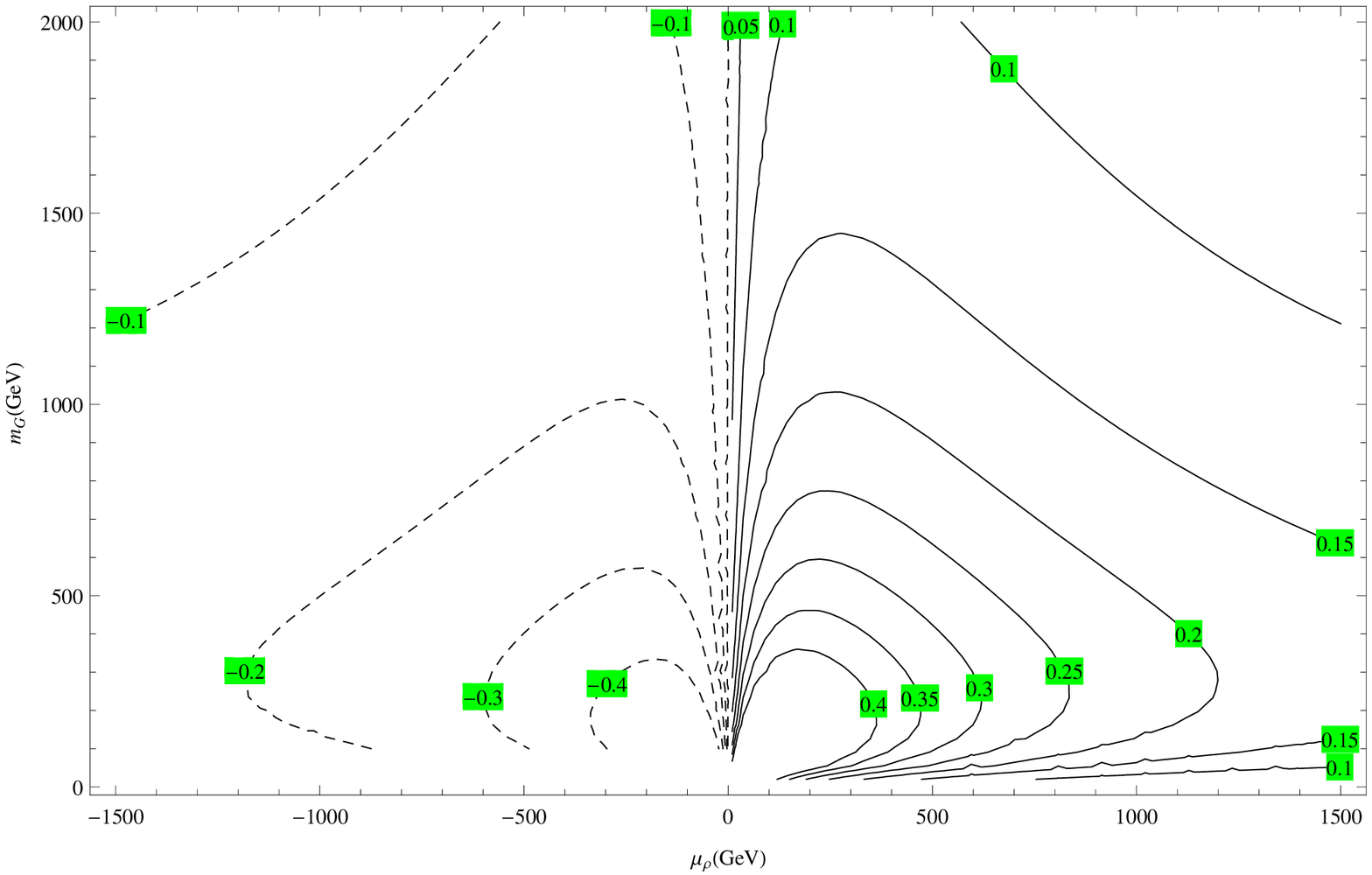}
\caption{$\Delta a_\mu$ plotted against $\mu_\rho$ and $m_G$\\
$m_{\tilde{l}_2}=m_{\tilde{l}_{L_2}}=m_{\tilde{l}_{R_2}}=m_{\tilde{\nu}_{L_2}}=m_{\tilde{\nu}_{R_2}}$,\\
 $m_{\tilde{l}_3}=m_{\tilde{l}_{L_3}}=m_{\tilde{l}_{R_3}}=m_{\tilde{\nu}_{L_3}}=m_{\tilde{\nu}_{R_3}}$,\\
    $\tan \gamma=60$, $m_{\tilde{l}_2} = 800$ GeV, $m_{\tilde{l}_3} = 2$ TeV,\\
     $m_B=m_\lambda=m_G$,\\
  $\theta_L=\theta_R=\frac{\pi}{4}$,
 $\theta_{\nu_L}=\theta_{\nu_R}=\frac{\pi}{4} $.   }%
    \label{Amu-mu-mG-34-b}
    \end{minipage}
\end{figure*}

Remember that the SUSYE331 contribution to the muon MDM  is proportional
 to $\tan \gamma$. The results given in Fig.\ref{AmutotalPlot} show that
 if the $\tan \gamma =60$, the interested $m_{SUSY}$ region is $[600, 800]$ GeV.
 We impose the upper bound for
 $\tan{\gamma}=60$ (ACHARD 04) \cite{PDG}.
 Hence, we numerically study the  SUSYE331 contribution to the muon MDM  in
 the case $\tan \gamma =60$ and the slepton mass hierarchy between the second
 and third family is retained. In  Fig.\ref{Amu-mu-mG-34-a},
  and Fig.\ref{Amu-mu-mG-34-b} we plot muon MDM on the
$m_G$ and $\mu_\rho$ plane with the same condition as above except the
mass of the generation is taken to be 500 GeV for  Fig. \ref{Amu-mu-mG-34-a}
and 800 GeV for Fig. \ref{Amu-mu-mG-34-b} respectively and the mass of
the third generation to be 2TeV. The result given in the Fig.\ref{Amu-mu-mG-34-a}
show that the upper bound for $m_G=1500$ GeV and $\mu_\rho =1500$ GeV.  However
the results given in the Fig.\ref{Amu-mu-mG-34-b}
set the upper bound for $m_G=1100$ GeV and $\mu_\rho=1200 $ GeV, respectively.

\begin{figure*}[!ht]\label{Amu-mL-mR}
\centering
\begin{minipage}{.5\textwidth}
  \centering
  \includegraphics[width=1.0\linewidth]{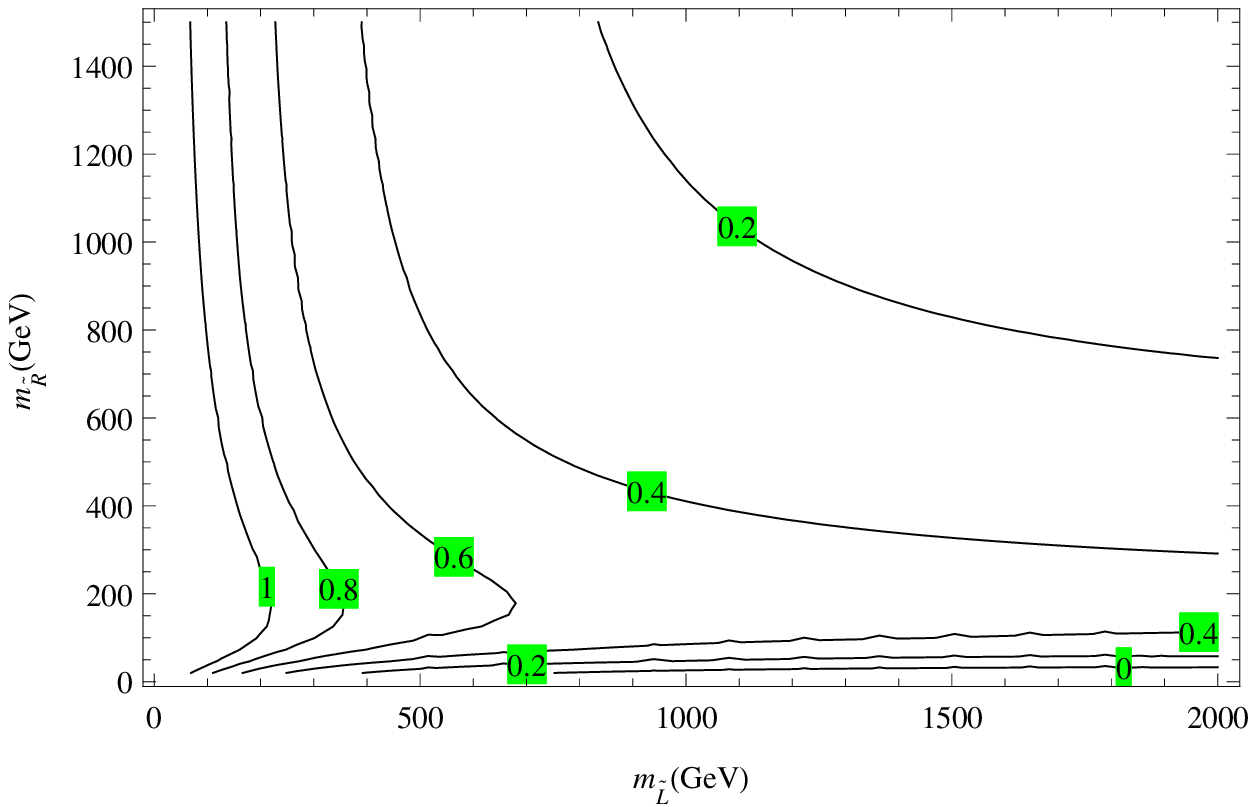}
    \caption{$\Delta a_\mu $ plot against $m_{\tilde{L}}$ and $m_{\tilde{R}}$.\\
    $m_{\tilde{l}_{L_2}}=m_{\tilde{\nu}_{L_2}}=m_{\tilde{l}_{L_3}}
=m_{\tilde{\nu}_{L_3}}=m_{\tilde{L}}$,\\
$m_{\tilde{l}_{R_2}}=m_{\tilde{\nu}_{R_2}}=m_{\tilde{l}_{R_3}}
=m_{\tilde{\nu}_{R_3}}= m_{\tilde{R}} $,\\
$\tan{\gamma}=60, \mu_\rho=140$ GeV, $m_\lambda$ =1 TeV }%
         \label{Amu-mL-mR-a}
\end{minipage}%
\begin{minipage}{.5\textwidth}
  \centering
  \includegraphics[width=1.0\linewidth]{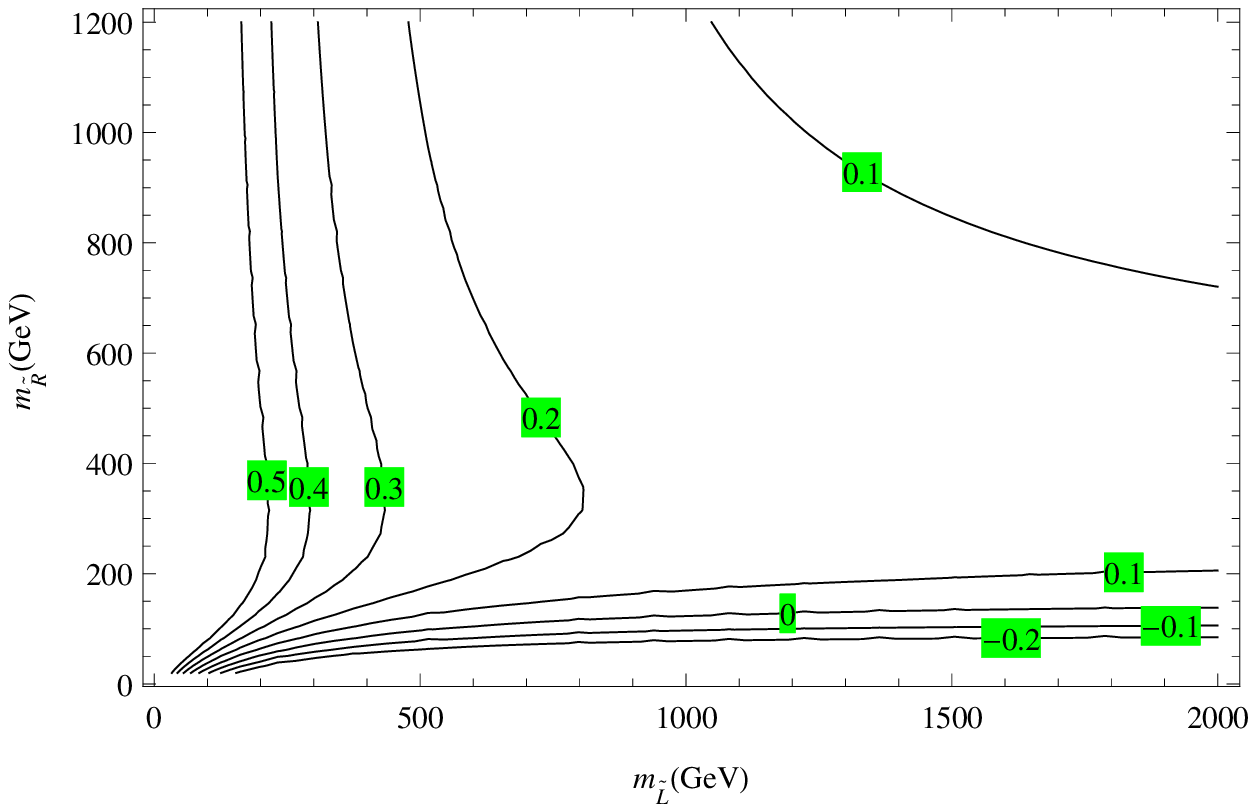}
    \caption{$\Delta a_\mu $ plot against $m_{\tilde{L}}$ and $m_{\tilde{R}}$.\\
    $m_{\tilde{l}_{L_2}}=m_{\tilde{\nu}_{L_2}}=m_{\tilde{l}_{L_3}}
=m_{\tilde{\nu}_{L_3}}=m_{\tilde{L}}$,\\
$m_{\tilde{l}_{R_2}}=m_{\tilde{\nu}_{R_2}}=m_{\tilde{l}_{R_3}}
=m_{\tilde{\nu}_{R_3}}= m_{\tilde{R}} $,\\
$\tan{\gamma}=60, \mu_\rho=140$ GeV, $m_\lambda$ =2 TeV    }%
         \label{Amu-mL-mR-b}
\end{minipage}
\end{figure*}

Next we will take into account the current upper bound of the mass
of Bino 350 GeV at the mass of top quark 174 GeV based on CP
violating phase \cite{PDG}. We have set
$m_{\tilde{l}_{L_2}}=m_{\tilde{\nu}_{L_2}}=m_{\tilde{l}_{L_3}}
=m_{\tilde{\nu}_{L_3}}=m_{\tilde{L}}$  and
$m_{\tilde{l}_{R_2}}=m_{\tilde{\nu}_{R_2}}=m_{\tilde{l}_{R_3}}
=m_{\tilde{\nu}_{R_3}}= m_{\tilde{R}} $ and $\tan{\gamma}=60,
\mu_\rho=140$ GeV. The mass of other gauginos $m_\lambda$ is set
to be 1 TeV (Fig.\ref{Amu-mL-mR-a}) and 2 TeV (Fig.\ref{Amu-mL-mR-b}). These figures illustrate the effects of
varying $m_{\tilde{L}}, m_{\tilde{R}}$ to the SUSYE331
contribution to the muon MDM. Combining the muon MDM from
experimental and the theoretical predicted given in the (\ref{Amu-mL-mR-a})
and (\ref{Amu-mL-mR-b}), we obtain the interested
region of the parameter space. Especially, the region of the
parameter space of  $m_{\tilde{R}}$ is very large $m_{\tilde{R}}>
20$ GeV while that of $m_{\tilde{L}}$ is slightly constrained and
depends on the value of gaugino mass. If we fixed  $m_\lambda =
1$ TeV the results given in Fig.\ref{Amu-mL-mR-a} show the lower
bound of $m_{\tilde{L}}\simeq 400$ GeV and
there is no upper bound for $m_{\tilde{R}}$.   From
Fig. \ref{Amu-mL-mR-b} we can find the upper bound mass of the
left slepton
$m_{\tilde{L}} \leq 800$ GeV.

\begin{figure*}[!ht]\label{Amu-mL2-mR2}
\centering
\begin{minipage}{.5\textwidth}
  \centering
  \includegraphics[width=1.0\linewidth]{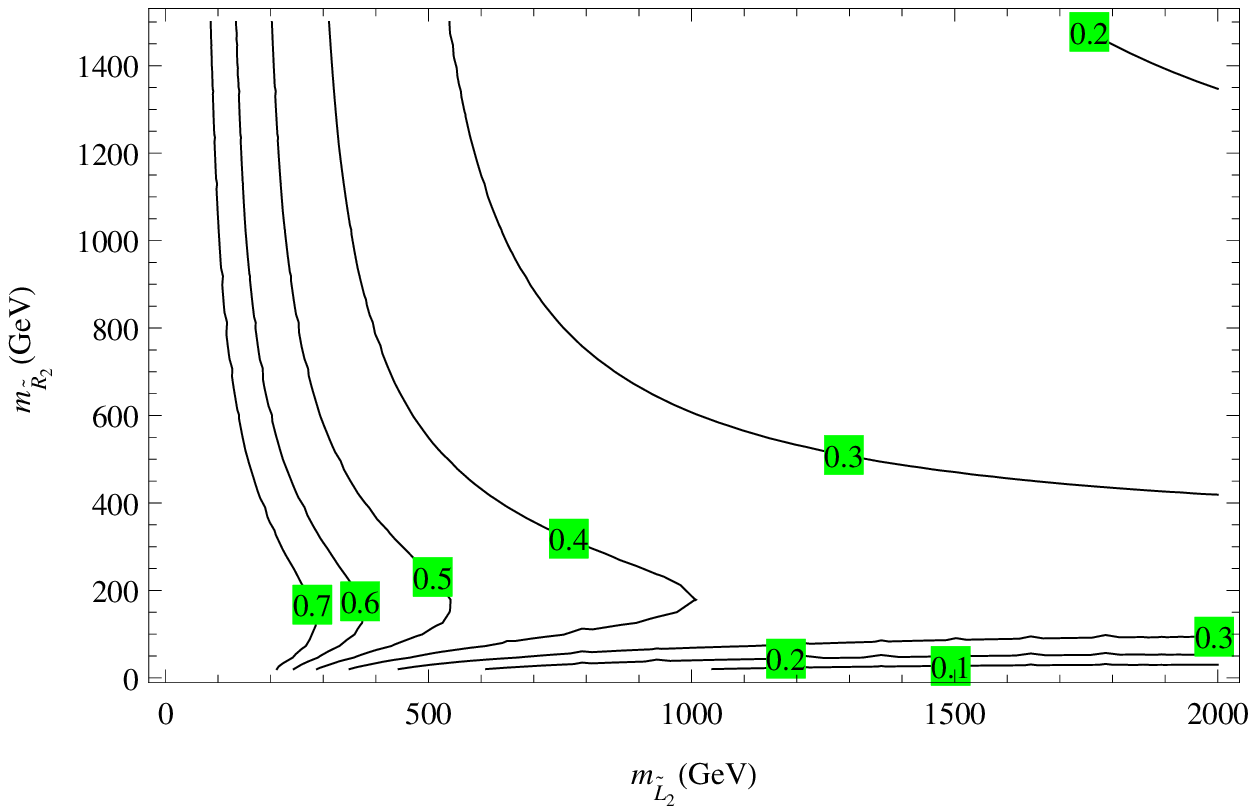}
    \caption{$\Delta a_\mu $ plot against $m_{\tilde{L}_2}$ and $m_{\tilde{R}_2}$.\\
    $m_{\tilde{l}_{L2}}=m_{\tilde{\nu}_{L2}}=m_{L_2}$ ,
$m_{\tilde{l}_{R2}}=m_{\tilde{\nu}_{R2}}=m_{R_2}$ \\
$m_{\tilde{l}_{L3}}=m_{\tilde{l}_{R3}}=m_{\tilde{\nu}_{L3}}
=m_{\tilde{\nu}_{R3}}=800 $ GeV \\
$\tan \gamma =60, \mu_\rho =140$ GeV,
$m_B= 350$ GeV}%
         \label{Amu-mL2-mR2-a}
\end{minipage}%
\begin{minipage}{.5\textwidth}
  \centering
  \includegraphics[width=1.0\linewidth]{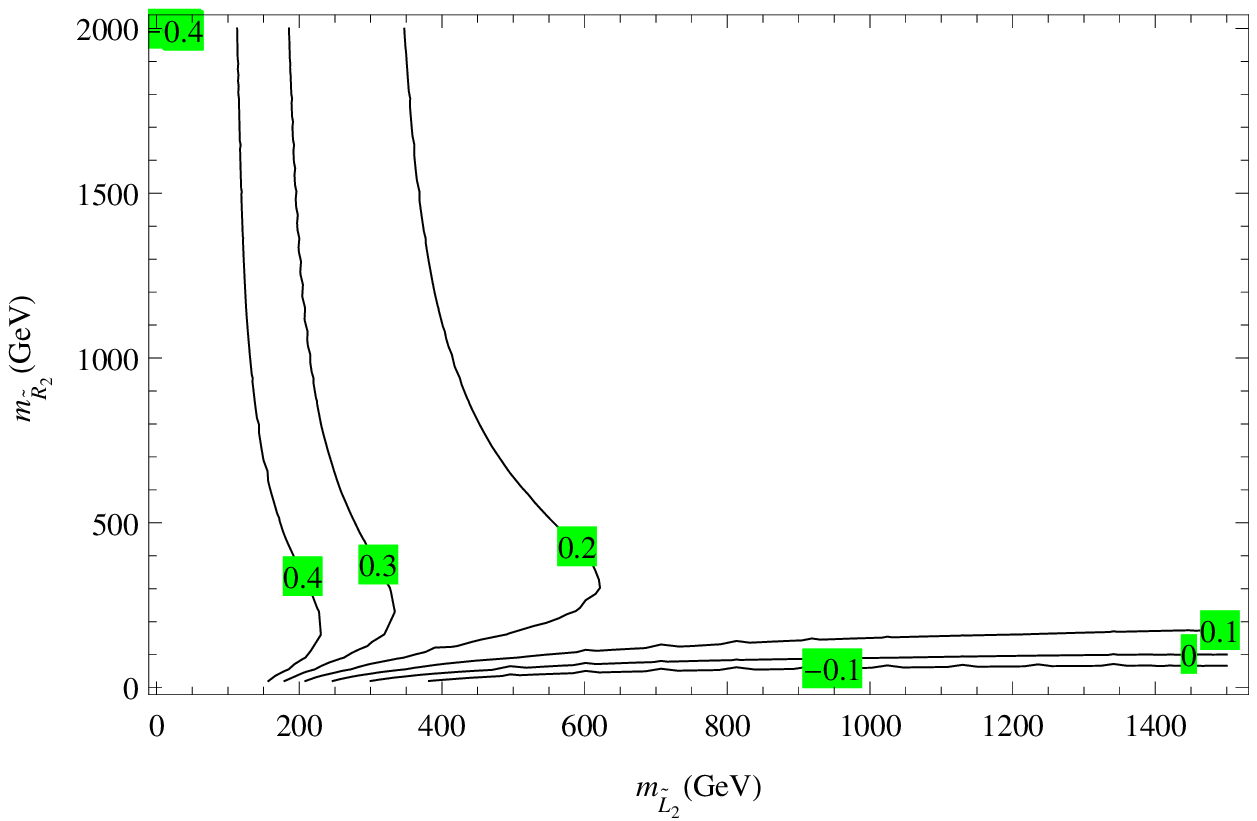}
    \caption{$\Delta a_\mu $ plot against $m_{\tilde{L}_2}$ and $m_{\tilde{R}_2}$.\\
    $m_{\tilde{l}_{L2}}=m_{\tilde{\nu}_{L2}}=m_{L_2}$ ,
$m_{\tilde{l}_{R2}}=m_{\tilde{\nu}_{R2}}=m_{R_2}$ \\
$m_{\tilde{l}_{L3}}=m_{\tilde{l}_{R3}}=m_{\tilde{\nu}_{L3}}
=m_{\tilde{\nu}_{R3}}=800 $ GeV \\
$\tan \gamma =60, \mu_\rho =140$ GeV,
$m_B= 350$ GeV}%
         \label{Amu-mL2-mR2-b}
\end{minipage}
\end{figure*}
Furthermore, by choosing the upper bound for the left-handed slepton mass above of third generation
$m_{\tilde{l}_{L3}}=m_{\tilde{l}_{R3}}=m_{\tilde{\nu}_{L3}}
=m_{\tilde{\nu}_{R3}}=800 $ GeV and fixing $\tan \gamma =60, \mu_\rho =140$ GeV,
$m_B= 350$ GeV, we plot the SUSYE331 contribution to the muon MDM on the plane
of $m_{\tilde{l}_{L2}}=m_{\tilde{\nu}_{L2}}=m_{L_2}$ ,
$m_{\tilde{l}_{R2}}=m_{\tilde{\nu}_{R2}}=m_{R_2}$ in Fig.\ref{Amu-mL2-mR2-a} and Fig.\ref{Amu-mL2-mR2-b}.
The results for the cases of $m_\lambda =1$ TeV and $m_\lambda=2$ TeV are shown in
Fig.\ref{Amu-mL2-mR2-a}   and Fig.\ref{Amu-mL2-mR2-b}, respectively.
Comparing with the experimental results, we find the lower bound of the
$m_{L_2} > 500$ GeV for fixing $m_\lambda = 1$ TeV and
the upper bound for the mass of
the left-handed slepton of the second generation
is 600 GeV for fixing $m_\lambda=2$ TeV. There is no bound of the right-handed
slepton mass of the second generation.

\begin{figure*}[!ht]\label{Amu-mnL2-mR2}
\centering
\begin{minipage}{.5\textwidth}
  \centering
  \includegraphics[width=1.0\linewidth]{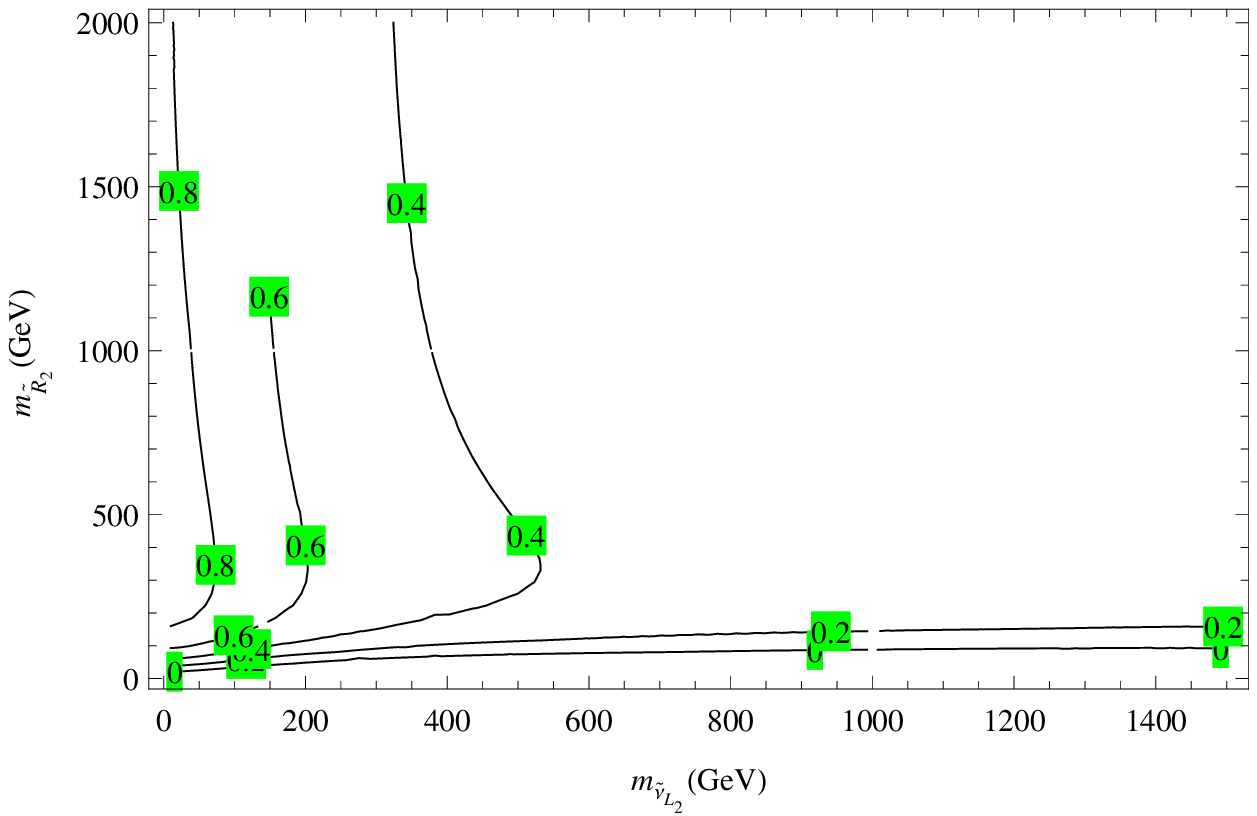}
    \caption{$\Delta a_\mu $ plot against $m_{\tilde{\nu}_{L_2}}$ and $m_{\tilde{R}_2}$. \\
    $\theta_R=\theta_L=0$ and $\theta_{\nu_R}=\theta_{\nu_L}
     =\frac{\pi}{4}$,  \\
    $\tan \gamma =60,$
     $\mu_\rho= 140$ GeV,\\
      $m_{R_3} = 800$ GeV, $m_{\tilde{L}_2}=600$ GeV,\\
       $m_B= 350$ GeV, $m_\lambda=1$ TeV  }%
\label{Amu-mnL2-mR2-a}
\end{minipage}%
\begin{minipage}{.5\textwidth}
  \centering
  \includegraphics[width=1.0\linewidth]{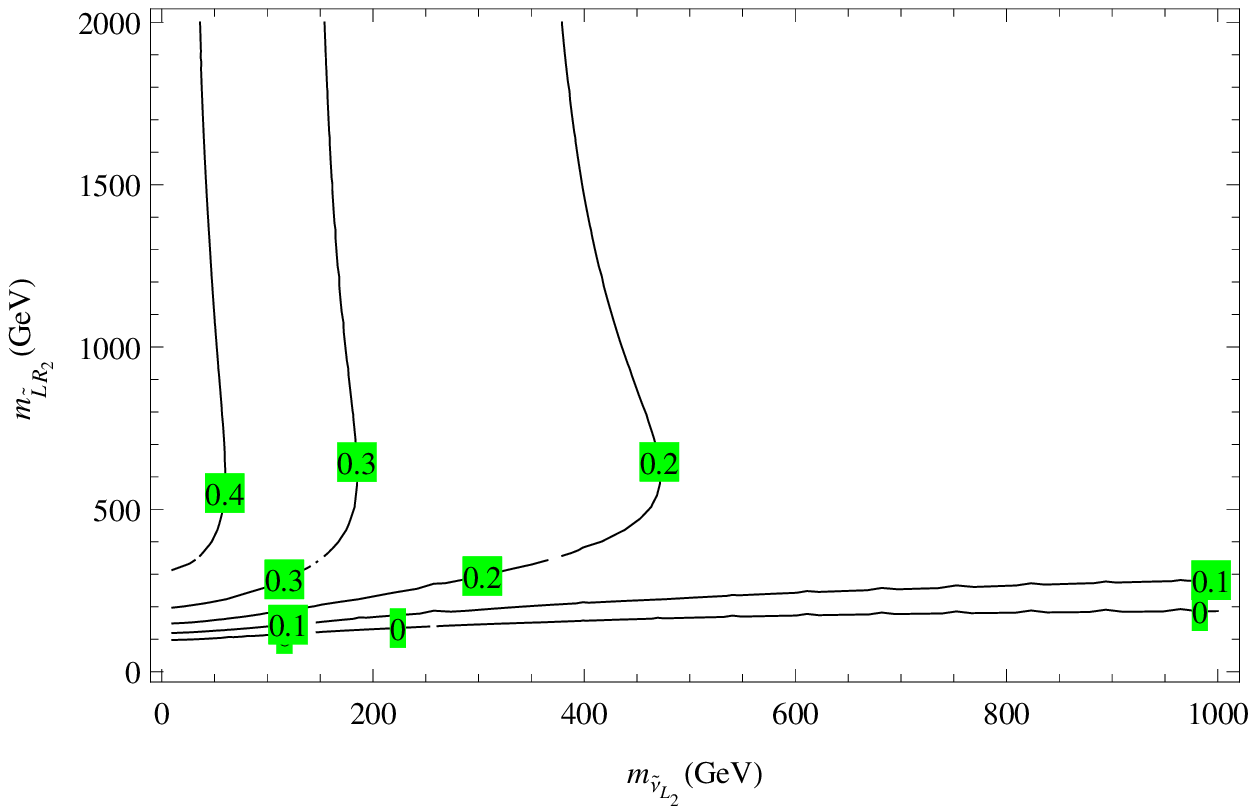}
\caption{$\Delta a_\mu $ plot against $m_{\tilde{\nu}_{L_2}}$ and $m_{\tilde{R}_2}$. \\
    $\theta_R=\theta_L=0$ and $\theta_{\nu_R}=\theta_{\nu_L}
     =\frac{\pi}{4}$,  \\
    $\tan \gamma =60,$
     $\mu_\rho= 140$ GeV, \\
     $m_{R_3} = 800$ GeV, $m_{\tilde{L}_2}=600$ GeV,\\
      $m_B= 350$ GeV, $m_\lambda=2$ TeV  }%
\label{Amu-mnL2-mR2-b}
\end{minipage}
\end{figure*}

    We have investigated the maximal mixing  case in which
     $ \theta_{\nu_R}=\theta_{\nu_L}=\frac{\pi}{4} $ are related
     to neutrino mixing and $ \theta_R=\theta_L=\frac{\pi}{4} $ are
     related to charge lepton mixing.  Next, we will investigate smaller
     case $\theta_R=\theta_L=0$ and $\theta_{\nu_R}=\theta_{\nu_L}
     =\frac{\pi}{4}$. In Figs. \ref{Amu-mnL2-mR2-a} and
     \ref{Amu-mnL2-mR2-b}
     we have plotted the MDM on the $ m_{\tilde{\nu}_{L_2}} $ and $m_{R_2}$
     where we have used the above constraint for
     $m_{\tilde{L}_2}=600$ GeV and other parameters are fixed as: $\tan \gamma =60,$
     $\mu_\rho= 140$ GeV, $m_{R_3} = 800$ GeV, $m_B= 350$ GeV,
     $m_\lambda=1$ TeV for the left -side  figure and $m_\lambda=2$ TeV
     for the right-side  figure.
     The mass of the sneutrino has  to be smaller than
     $550$ GeV to address $2-3.6$  $\sigma$ discrepancy if $m_\lambda =2$ TeV.
     The lower bound of the right-handed slepton mass of the second generation
     is around tens of GeV  for fixing $m_\lambda =1$ TeV  and is
     a hundred GeV for fixing $m_\lambda =2$ TeV.

\section{Conclusions}
In this paper we have examined in detail the muon $g-2$ in the frame
work of the  SUSYE331 model. We calculated one-loop SUSYE331
contributions to the muon MDM based on both the mass eigenstate
and the weak eigenstate methods. The mass eigenstates of
the neutralino and chargino  are obtained using approximation method.
To recognize the effects of the SUSYE331
parameters to the muon MDM , we work with the
analytical expressions of the muon MDM  based on the mass
insertion method. We have considered all parameters of SUSYE331 as
free parameters. In our calculation we have made some assumptions:
maximal mixing, mixing terms $A_\tau, A_\mu $, and $A^{L,R}_{\mu
\tau}$ are small and are neglected in our calculation. In
particularly, by taking the limit $\mu_\rho, m_\lambda,
m_{\tilde{L}}, m_B$ to $m_{SUSY}$ we obtain the reduced analytical
expressions of the contribution of the SUSYE331 to the muon MDM.
Results show that the SUSYE331 contribution to the muon MDM  is
enhanced in the small region of the $m_{SUSY}$  and the large
values of the  $\tan \gamma$. We have investigated for both small
and large values of ratio of two vacua $\tan{\gamma}$. The numerical
results show that in order to consist with the experimental bound of
the muon MDM , the  $m_{SUSY} \simeq75$ GeV  for $\tan \gamma
=1$ and $m_{SUSY}=900$ for $\tan \gamma =60$. On the other hand, we also investigated the
SUSYE33 contribution to the muon MDM in the case of
the mass hierarchy between the second and the third generation.
In the case which
$\tan{\gamma}$ is small ($\tan{\gamma}=5$) and one generation of the slepton mass
is fixed at the ${\cal O}(1)$ TeV,
the light slepton
particle mass is bounded from  $100$ GeV to $200$ GeV and $\Delta a^\mu$ can be
comparable with the current limit on the muon MDM. When
$\tan{\gamma}$ is large ($\tan{\gamma}=60$) the mass of the light left-handed
slepton particle is bound to 800 GeV .
Finally, we would like to comment on the case with
the maximal flavor mixing only in the sneutrino sector, we obtain the upper bound of the mass of sneutrino is
 is  550 GeV. These values can be examined at LHC or
furure collider  ILC.

\section{Feynman diagrams}

\begin{acknowledgments}
 D. T. Binh thanks Le Tho Hue for his suggestion. This research is funded by Vietnam National Foundation
  for Science and Technology Development (NAFOSTED) under grant number 103.01-2014.51.
\end{acknowledgments}

\section{Appendix section}\label{app}

In this section we will define some integrals  used in our calculation.
The following integral is defined in \cite{Dadychev}.
\bea
I^{(N)}({\nu_i};{\nu_j})&\equiv& \int \frac{d^Dk}{(k^2-m_1^2)^{\nu_1}...(k^2-m_N^2)^{\nu_N}} \crn
&=& \int \frac{d^Dk}{\Pi_{j=1}^N(k^2-m_j^2)^{\nu_j}}
\nn \eea
In special case $D=4$ and there are two masses,  we have:
\bea
I^{(2)}(\nu_1,\nu_2; m_1,m_2)&=& \pi^{2}i^{-3}(-m_2)^{2-\nu_1-\nu_2}\frac{\Gamma(\nu_1+\nu_2-2)}{\Gamma(\nu_1+\nu_2)}  \crn
&\times& _2F_1(\nu_1+\nu_2-2,\nu_1;\nu_1+\nu_2;1-\frac{m_1^2}{m_2^2})\nn
\eea
The loop integrals are defined as :
\[
I_N(m_1^2,...,m_N^2)=\frac{i}{\pi^2}\int \frac{d^4k}{(k_1^2-m_1^2)...(k_1^2-m_N^2)},
\]
\[
J_N(m_1^2,...,m_N^2)=\frac{i}{\pi^2}\int \frac{k^2 d^4k}{(k_1^2-m_1^2)...(k_1^2-m_N^2)}.
\]


\begin{thebibliography}{9}
\bibitem{PDG} J. Beringer et al, (Particle Data Group), Phys. Rev. D  {\bf 86}, 010001 (2012).
\bibitem{331ver1} H. N. Long, Phys. Rev. D {\bf 54}  (1996) 4691.
\bibitem{331ver2} F. Pisano, V. Pleitez, Phys. Rev. D 46 (1992) 410,
\bibitem{331ver3} R. Foot, H. N. Long, T. A. Tran, Phys. Rev. D {\bf 50}  (1994) R34.
\bibitem{331ver4} C. A. de S. Pires, P. S. Rodrigues da Silva,
JCAP 0712 (2007) {\bf 012}.
\bibitem{331ver5} J. K. Mizukoshi,
 C. A. de S. Pires, F. S. Queiroz, P. S. R. da Silva,
 Phys. Rev. D 83 (2011) 065024.
\bibitem{331ver6} F. Queiroz, C. A. de S. Pires, P. S. R.
da Silva, Phys. Rev. D {\bf 82}  (2010) 065018.
\bibitem{331ver7} D.~Cogollo, A.~V.~de Andrade, F.~S.~Queiroz and
P.~Rebello Teles,
  Eur.\ Phys.\ J.\ C {\bf 72}, 2029 (2012).
\bibitem{331ver8} J.~D.~Ruiz-Alvarez, C.~A.~de S.Pires, F.~S.~Queiroz,
D.~Restrepo and P.~S.~Rodrigues da Silva, Phys.\ Rev.\ D {\bf 86}, 075011 (2012).
\bibitem{331ver9} P.~V.~Dong, H.~N.~Long and H.~T.~Hung, Phys.
\ Rev.\ D {\bf 86}, 033002 (2012)
\bibitem{331ver10} A.~Alves, E.~Ramirez Barreto, A.~G.~Dias,
 C.~A.~de S.Pires, F.~S.~Queiroz and P.~S.~Rodrigues da Silva,
 Eur.\ Phys.\ J.\ C {\bf 73}, 2288 (2013).
\bibitem{331ver11} P. V. Dong, H. T. Hung, T. D. Tham , Phys.
 Rev. D {\bf 87}  (2013) 115003.
\bibitem{331ver12} P. V. Dong, T. Phong Nguyen, D. V. Soa,
Phys. Rev. D {\bf 88}, 095014 (2013).
\bibitem{331verE} P.  V.  Dong, H. N. Long, D. T. Nhung and
D. V.  Soa, Phys. Rev. D 73, 035004 (2006). 
\bibitem{Lepto} D. T. Huong,  P. V. Dong,  C. S.  Kim, N. T. Thuy,
 Phys. Rev. D, 91, 055023 (2015); P.  V Dong, D.  T.  Huong,
 Farinaldo S.  Queiroz, N. T. Thuy,  Phys. Rev. D 90, 075021 (2014)
\bibitem{g2muon331v1} C.A. de S. Pires, P. S. Rodrigues da Silva,
Phys. Rev. D 64 (2001) 117701. 
\bibitem{g2muon331v2} C.A. De Sousa Pires, P. S. Rodrigues da Silva,
Phys. Rev. D 65 (2002) 076011. 
\bibitem{g2muon331v3} N. A.  Ky, H. N. Long, D. V. Soa, Phys. Lett. B 486 (2000) 140. 
\bibitem{g2muon331v4}C.  Kelso, P. R. D. Pinheiro, F.  S. Queiroz,
W.  Shepherd. Eur. Phys. J. C 74 (2014) 2808.
\bibitem{queroslong} Chris Kelso, H. N. Long, R. Martinez, Farinaldo S. Queiroz,  \emph{Phys. Rev.}
\textbf{D 90}, 113011 (2014). 
\bibitem{SUSYnaturalness} M.J.G. Veltman, Acta Phys. Pol. B12(1981)437.
\bibitem{SUSYUnify} P. Langacker, M. Luo, Phys. Rev. D44 (1991) 817.
\bibitem{MSSMg-2} P. Fayet,
in Unification of the Fundamental Particles Interactions, edited by S. Fer-
rara, J. Ellis and P. van Nierwenhuizen (Plenum, New York, 1980) p. 587;
J.A. Grifols and A. Mendez, Phys. Rev. D26 (1982) 1809;
J. Ellis, J. S. Hagelin and D. V. Nanopoulos, Phys. Lett. B166 (1982) 283;
R. Barbieri and L. Maiani, Phys. Lett. B117 (1982) 203;
D. A. Kosower, L. M. Krauss and N. Sakai, Phys. Lett. B133 (1983) 305;
T. C. Yuan, R. Arnowitt, A. H. Chamseddine and P. Nath, Z. Phys. C26 (1984) 407;
J. C. Romao, A. Barroso, M. C. Bento and G. C. Branco, Nucl. Phys. B 250 (1985)
295; J. Lopez, D.V. Nanopoulos and X. Wang, Phys. Rev. D 49 (1991) 366; U. Chattopadhyay and P. Nath;
T.  Moroi, Phys. Rev. D {\bf 53} , 6565 (1996).
\bibitem{MSSM}J. Ellis and D. V. Nanopoulos, Phys. Lett. 110B (1982) 44;
I-H. Lee, Phys. Lett. {\bf  B 138}  (1984) 121; Nucl. Phys. {\bf B 246}  (1984) 120.
\bibitem{LFV}A. Brignole, A. Rossi, Nucl. Phys. B 701 (2004),3-53;
K. S. Babu and C. Kolda, Phys. Rev. Lett. 89 (2002) 241802; A.  Abada, D.  Das and C.  Weiland, JHEP
1203 (2012) 100; A.  Abada., D.  Das, A.  Vicent and C.
Weiland, JHEP 1209 (2012) 015.
\bibitem{SUSYE331} P. V. Dong, D. T. Huong, M. C. Rodriguez,
 H. N. Long,  Nucl. Phys.  {\bf B 772}, 150   (2007).
\bibitem{LFV1} P. T. Giang, L. T. Hue, D. T. Huong, H. N. Long,
 Nucl.  Phys.   {\bf  B 864} (2012) 85. 
\bibitem{LFV2} L. T. Hue, D. T. Huong, H. N. Long,
 Nucl. Phys. {\bf B 873} (2013) 207. 
\bibitem{chargedlepton} P. V. Dong, Tr. T.Huong, N. T. Thuy, H. N. Long,
  {\it JHEP} {\bf073},(2007) 0711.
\bibitem{gaugino}D. T. Huong, H. N. Long,
 {\it JHEP}, {\bf 049},(2008) 0807. 
\bibitem{g2SM}M. Davier. et al., Eur. Phys. J. C 66 (2010)127;
A. Hoecker, B. Malaescu, C. Z. Yuan and Z. Zhang, Eur. Phys. C 66 (2010) 1.
\bibitem{g2E821}Muon G-2 Collaboration, Phys. Rev. D 73 (2006).
\bibitem{Czarnecki}A.  Czarnecki and W. J. Marciano,
Phys. Rev. D 64 (2001) 013014.
\bibitem{Dadychev} A. I. Davydychev,
 J. Math. Phys. {\bf 32}  (1991) 4299.
\end{thebibliography}
\end{document}